\documentclass[aps,prd,twocolumn,showpacs,eqsecnum,nofootinbib,floatfix]{revtex4-1}
\pdfoutput=1

\usepackage{amssymb}
\usepackage{graphicx}
\usepackage{amsmath}
\usepackage{url}
\usepackage{rotating}
\usepackage{dcolumn}
\usepackage{url}
\usepackage{graphicx,color}

\font\FermiSmallfont=cmssq8 scaled 1200

\def\UMDppthead#1#2#3{
\null 
\begin{center}\vskip -1.0truein{\hbox to 7.5truen {
\hfill 
\vbox to 1in {\vfill \FermiSmallfont
              \hbox{#1}
              \hbox{#2}
              \hbox{#3}
              \vfill}
}}\vskip-0.0truein\end{center}}

\def\ts{\rm TS}
\def\apjs{Astrophys.\ J.\ Supp.}

\begin{document}

\title{Astrophysical and dark matter interpretations of extended gamma-ray emission from the Galactic Center}

\author{Kevork N.\ Abazajian} \email{kevork@uci.edu}
\author{Nicolas Canac} \email{ncanac@uci.edu}
\author{Shunsaku Horiuchi} \email{s.horiuchi@uci.edu}
\author{Manoj Kaplinghat} \email{mkapling@uci.edu}
\affiliation{Center for Cosmology, Department of Physics and Astronomy, University of
  California, Irvine, Irvine, California 92697 USA}


\begin{abstract}
We construct empirical models of the diffuse gamma-ray background
toward the Galactic Center. Including all known point sources and a
template of emission associated with interactions of cosmic rays with
molecular gas, we show that the extended emission observed previously in 
the Fermi Large Area Telescope data toward the Galactic Center is detected at
high significance for all permutations of the diffuse model
components. However, we find that the fluxes and spectra of the
sources in our model change significantly depending on the 
background model. In particular, the spectrum of the central Sgr
A$^\ast$ source is less steep than in previous works and the recovered
spectrum of the extended emission has large systematic uncertainties, especially
at lower energies. If the extended
emission is interpreted to be due to dark matter annihilation, we find
annihilation into pure $b$-quark and $\tau$-lepton channels to be
statistically equivalent goodness of fits. In the case of the pure
$b$-quark channel, we find a dark matter mass of
$39.4\left(^{+3.7}_{-2.9}\rm\ stat.\right)\left(\pm
7.9\rm\ sys.\right)\rm\ GeV$, while a pure $\tau^{+} \tau^{-}$-channel
case has an estimated dark matter mass of
$9.43\left(^{+0.63}_{-0.52}\rm\ stat.\right)(\pm 1.2\rm\ sys.)\ GeV$.
Alternatively, if the extended emission is interpreted to be
astrophysical in origin such as due to unresolved millisecond pulsars, 
we obtain strong bounds on dark matter annihilation, although systematic 
uncertainties due to the dependence on the background models are significant. 
\end{abstract}

\maketitle

\section{Introduction}

The Milky Way's Galactic Center (GC) harbors an extremely dense
astrophysical environment, with thousands of high-energy sources
detected in the X-ray within the inner
0.3$^\circ$~\cite{Muno:2008qy}, as well as numerous gamma-ray
emitting point sources~\cite{Nolan2012}. In addition, the GC is
expected to harbor high densities of dark matter (DM) with a
power-law increase in density toward the center, leading it
to be among the best places in which to find signals of DM 
particle annihilation or decay \cite{Springel:2008by}.
A leading candidate for cosmological dark matter is a
thermally produced weakly interacting massive particle (WIMP) that can
arise in many extensions of the Standard Model of particle
physics, whose annihilation is related to their production in the
early Universe~\cite{Feng:2010gw}. 

Several groups have found strong evidence for extended emission in the
gamma ray from the GC using data from the Large Area Telescope (LAT)
aboard the {\it Fermi Gamma-ray Space Telescope}. It has been shown
that the extended emission is consistent with the spatial profile
expected in DM halo formation simulations, the flux is consistent with
the annihilation rate of thermally produced WIMP DM, and the spectrum
is consistent with the gamma rays produced in the annihilation of
$\sim\!\! 10 - 30$ GeV DM to quarks or
leptons~\cite{Goodenough:2009gk,Hooper:2010mq,Hooper:2011ti,Boyarsky:2010dr,Abazajian:2012pn,Gordon:2013vta,Macias:2013vya}.
This triple consistency of the gamma-ray extended-source signal in the
GC with the WIMP paradigm has generated significant interest. In
addition, there are claims of signals consistent with the DM origin
interpretation in the ``inner Galaxy'' \cite{Hooper:2013rwa}, and in
unassociated point sources \cite{Berlin:2013dva}. The required 
dark matter mass and annihilation cross section is consistent with the 
constraints from Milky Way dwarf galaxies~\cite{Ackermann:2013yva}.

\begin{figure*}[h!t!]
\begin{sideways}
\makebox[1.6truein][c]{$\quad$ Diffuse Sources' Residual}
\end{sideways}
\includegraphics[width=1.68truein]{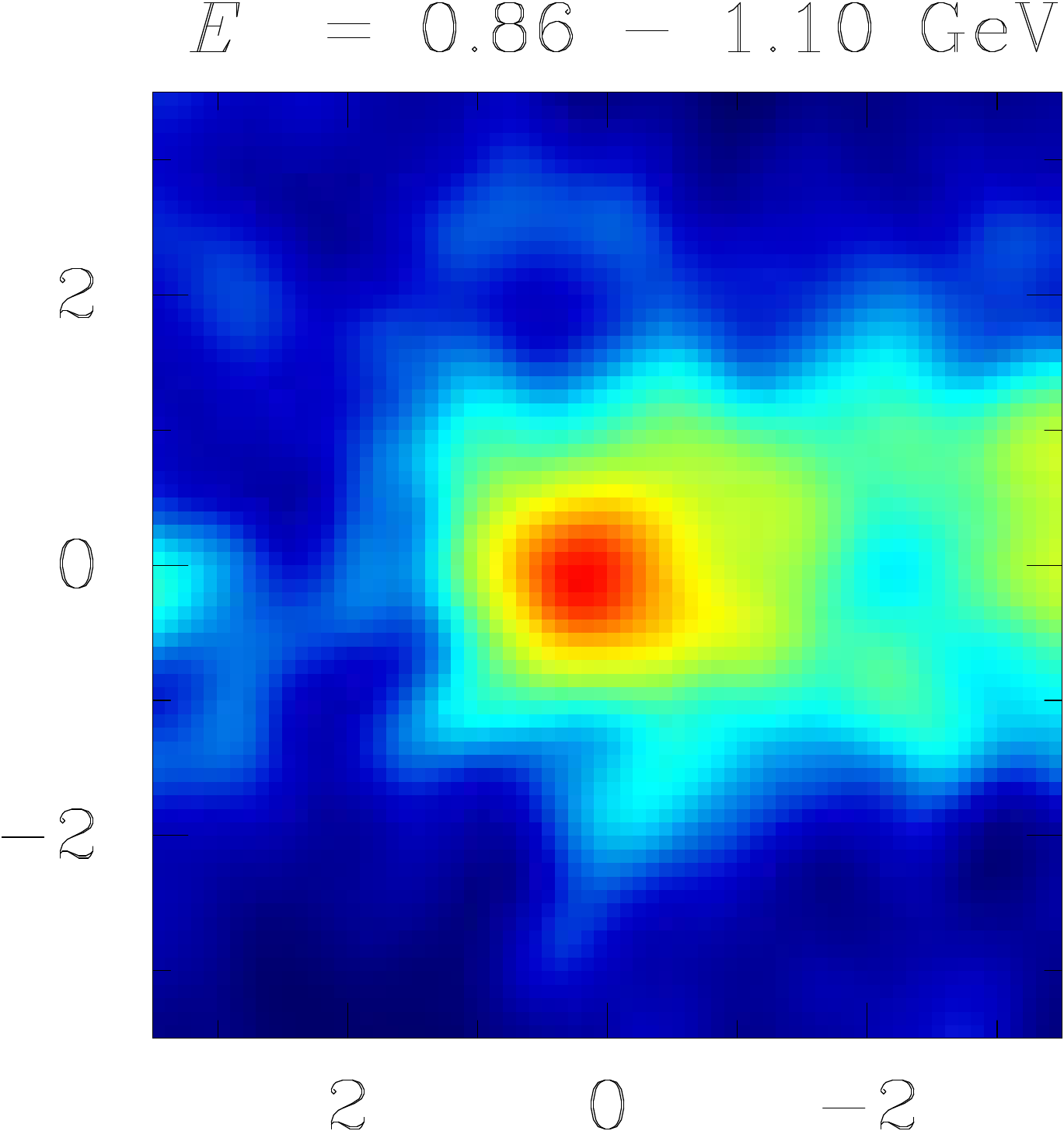}
\includegraphics[width=1.68truein]{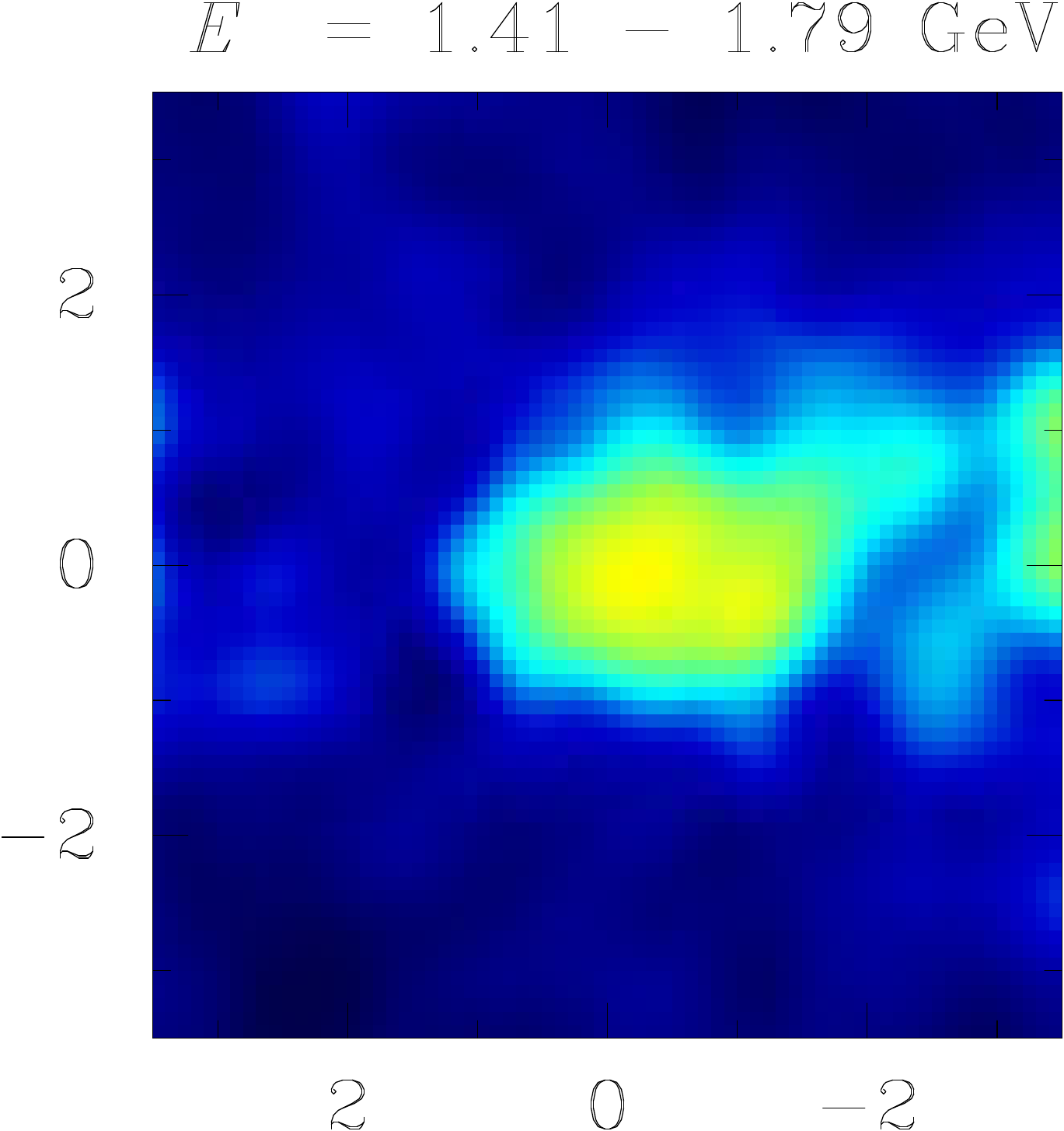}
\includegraphics[width=1.68truein]{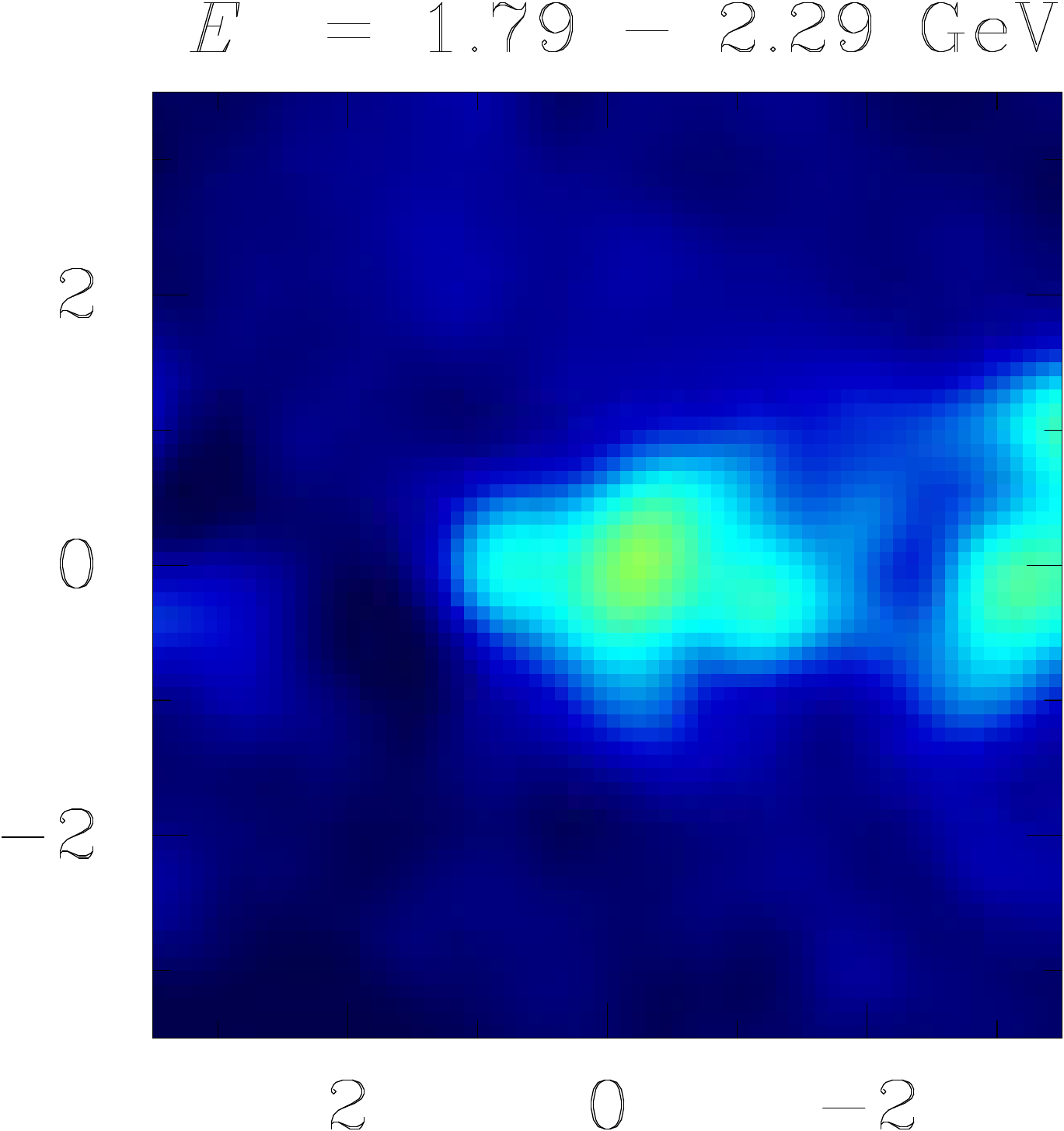}
\includegraphics[width=1.68truein]{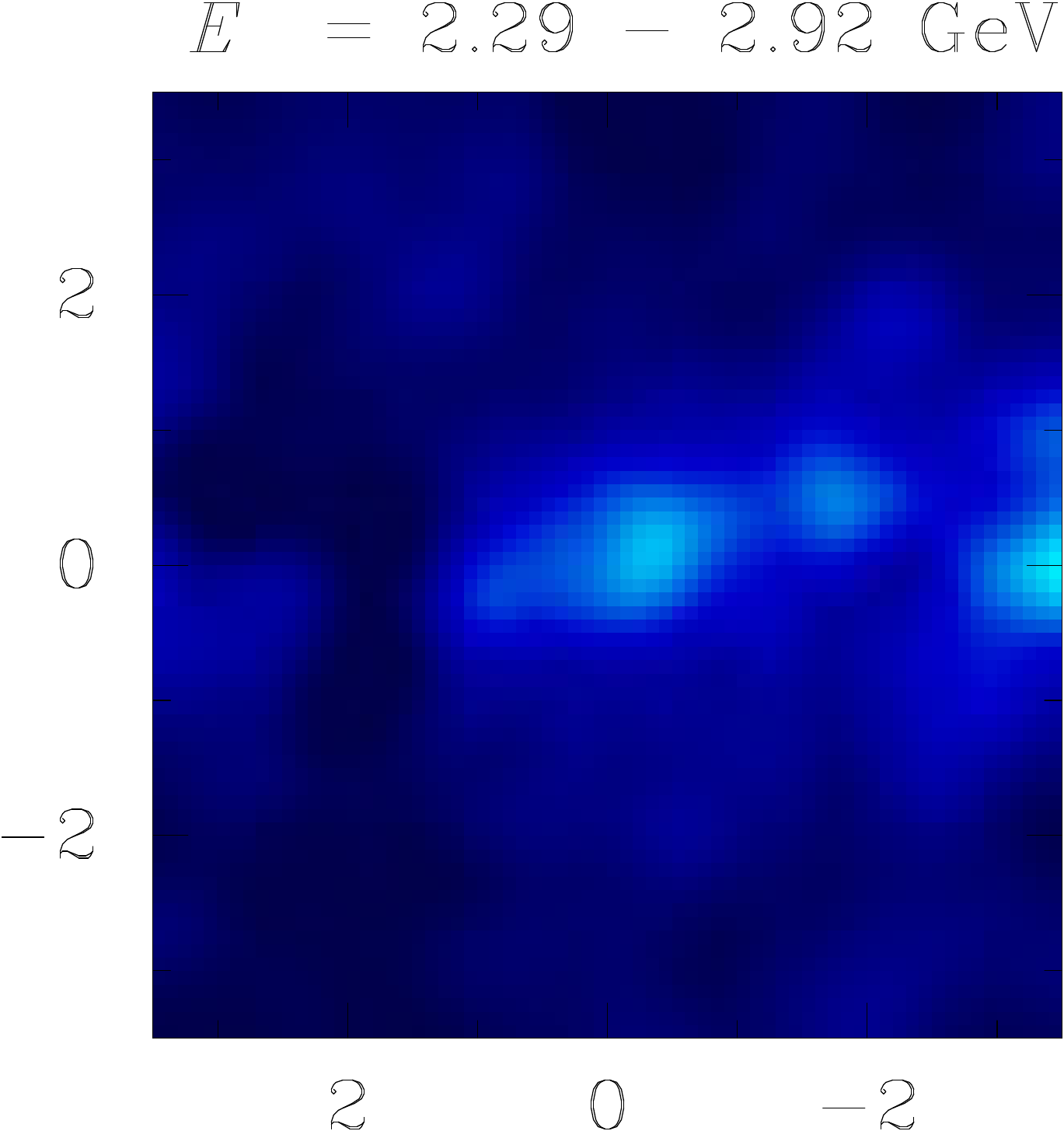}\\
\begin{sideways}
\makebox[1.6truein][c]{$\quad$ Diffuse Models}
\end{sideways}
\includegraphics[width=1.68truein]{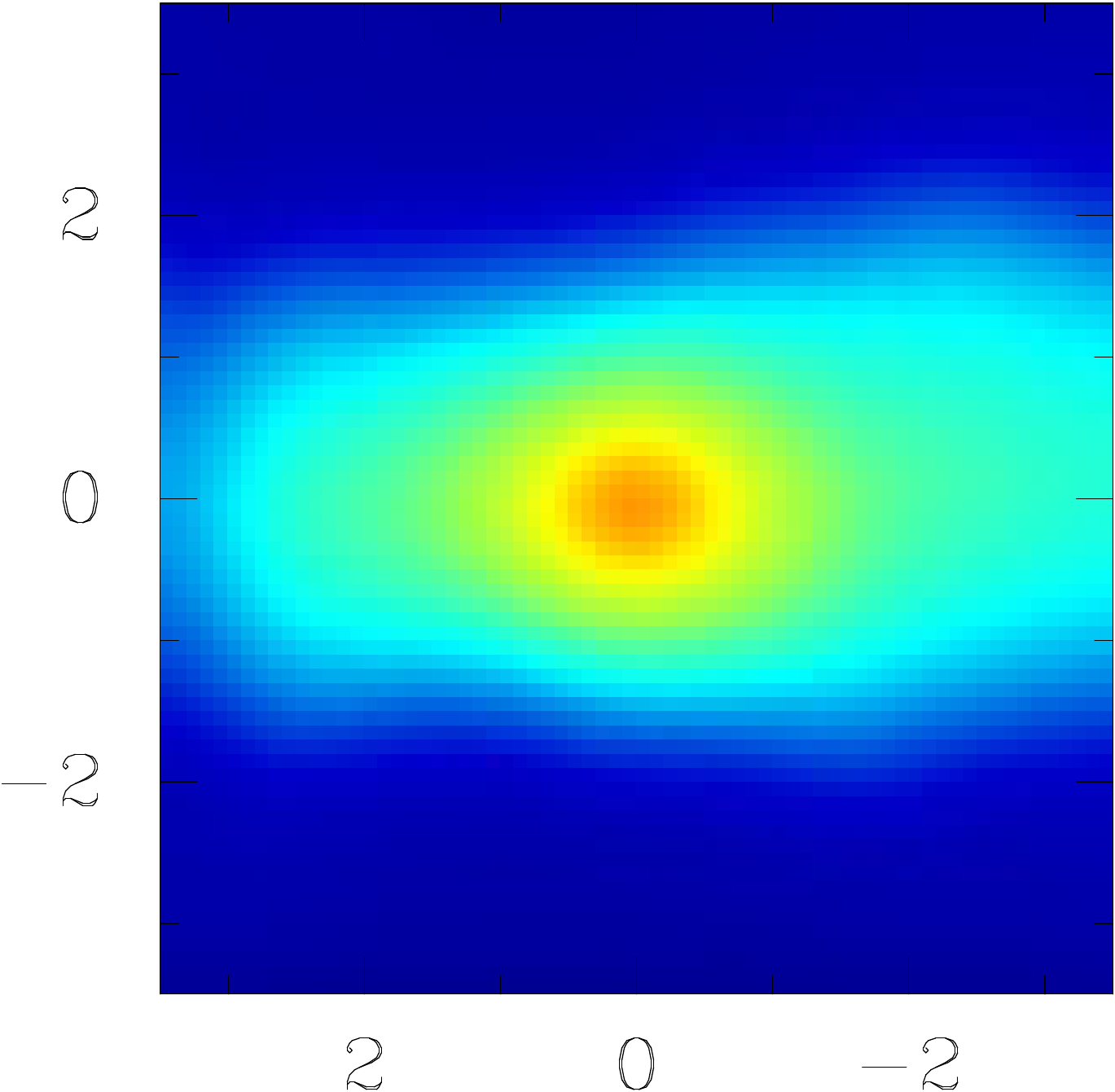}
\includegraphics[width=1.68truein]{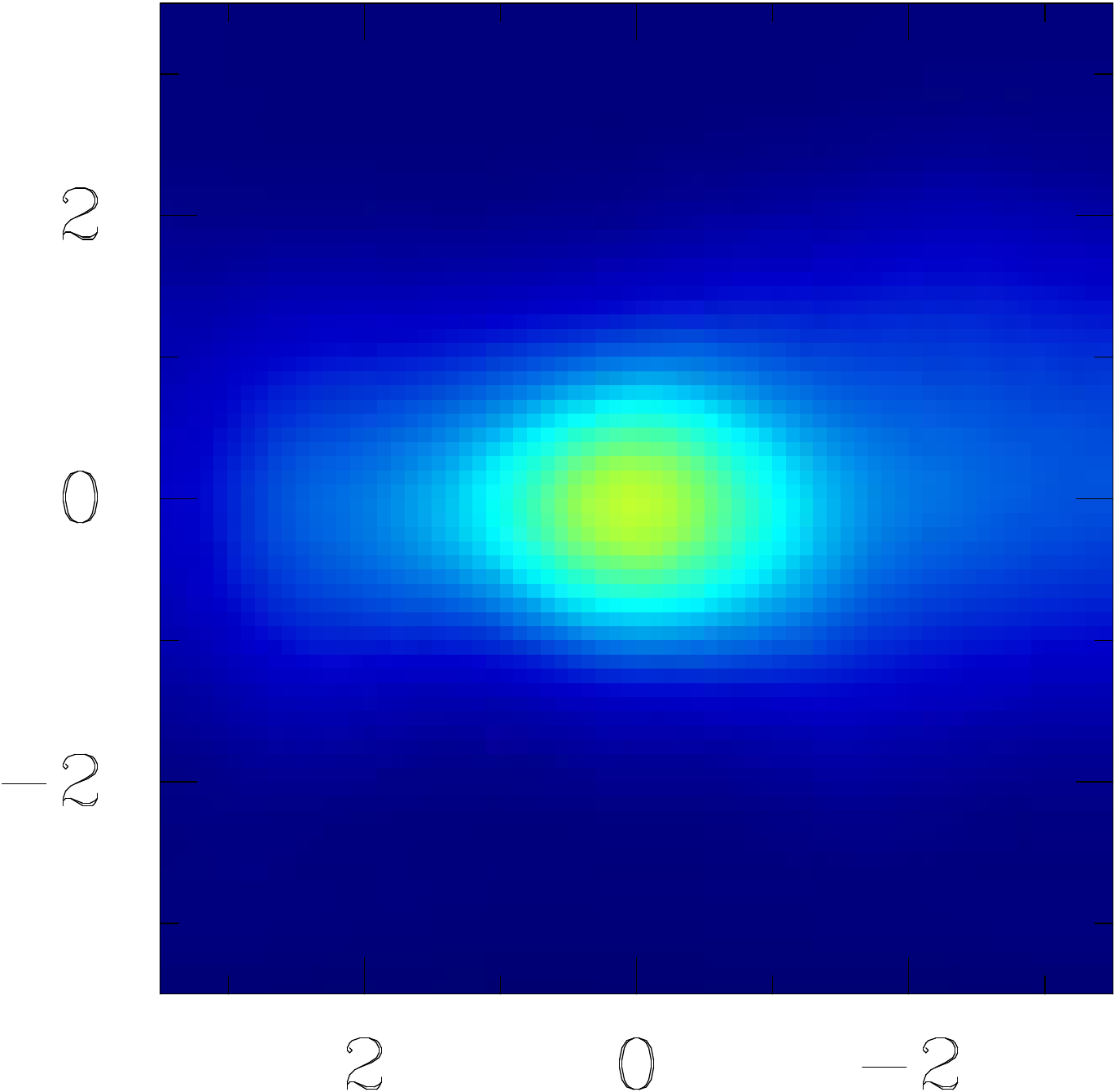}
\includegraphics[width=1.68truein]{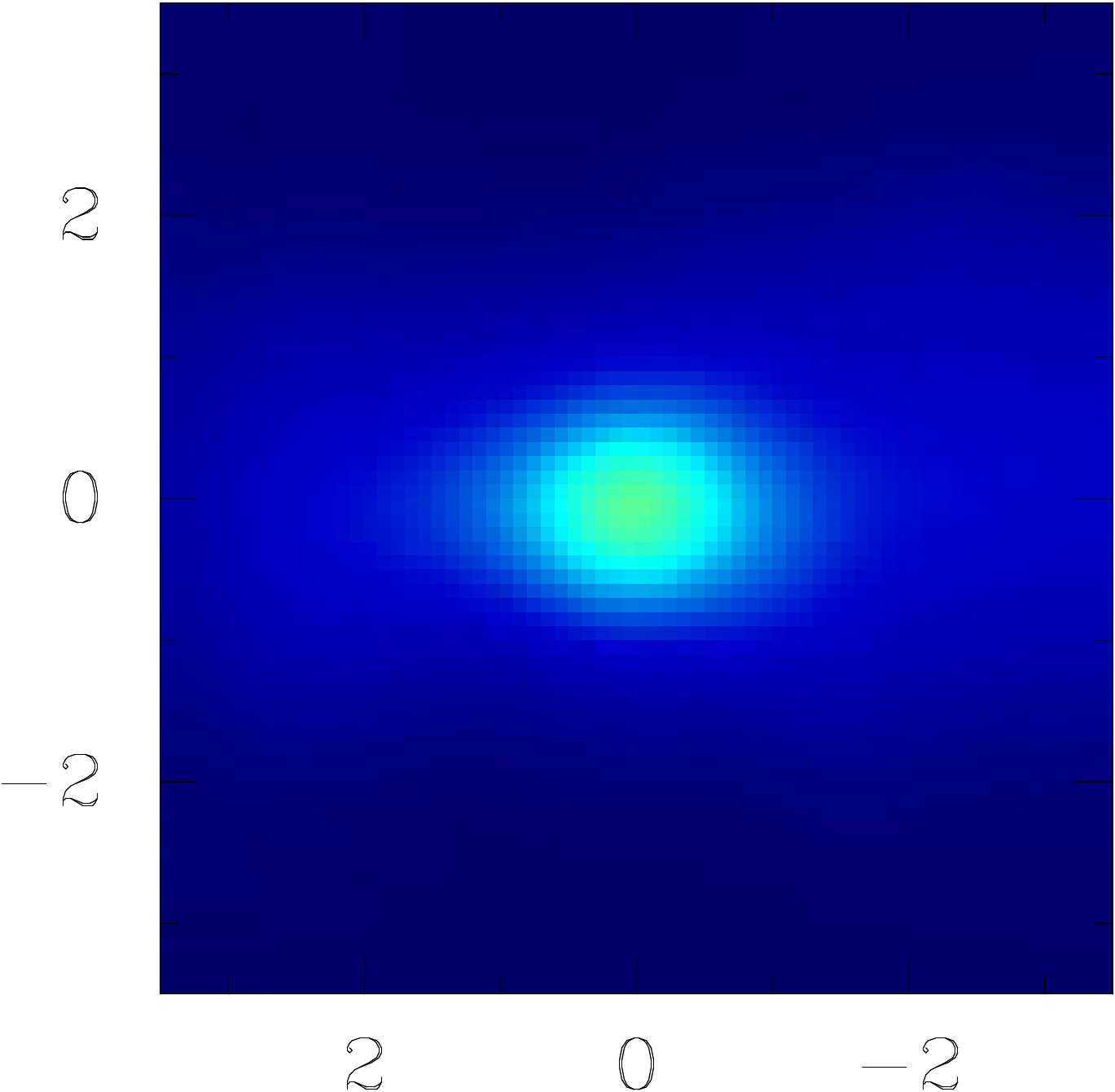}
\includegraphics[width=1.68truein]{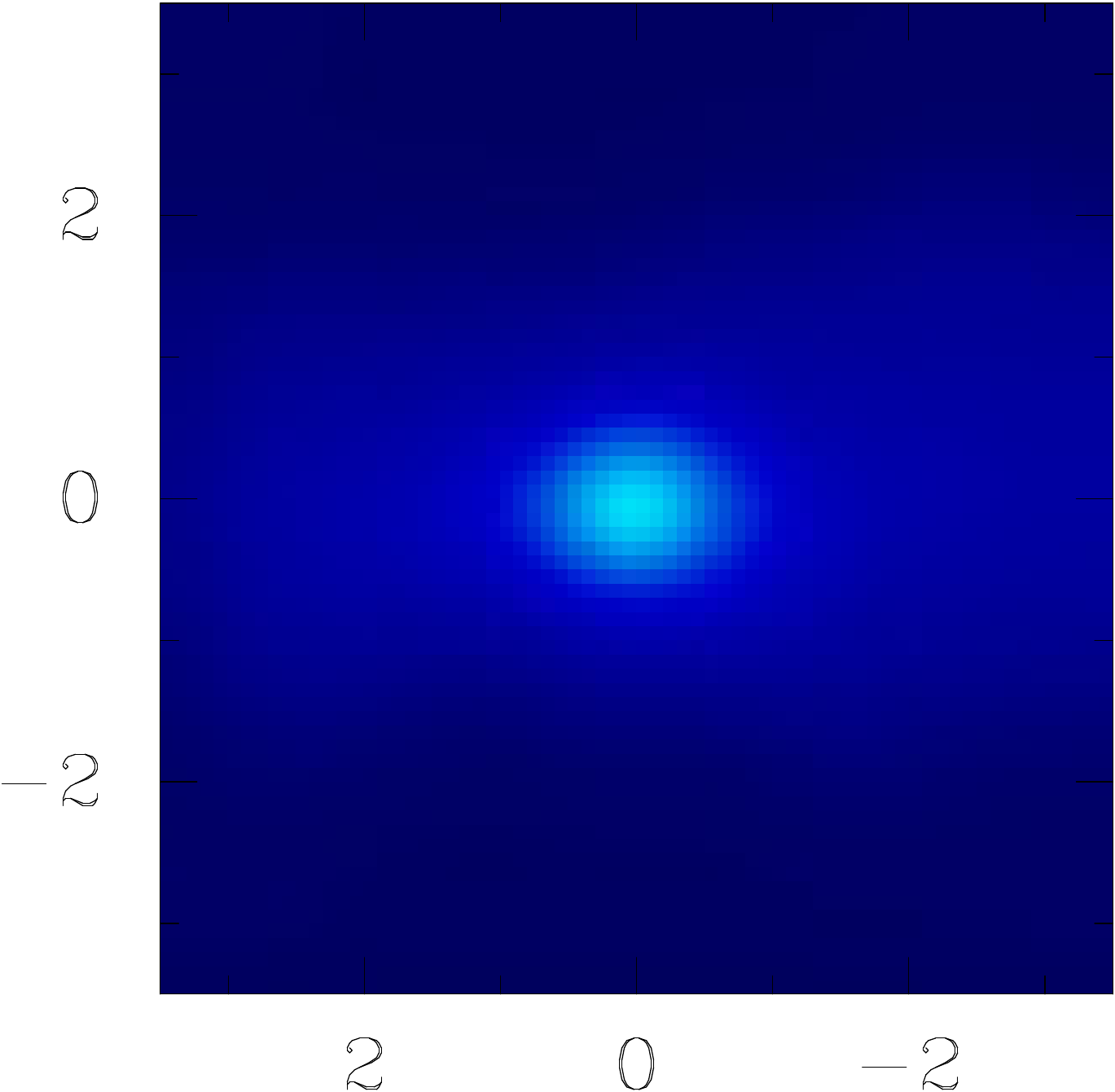}\\
\begin{sideways}
  \makebox[2.1truein][c]{\ $\qquad$\ \ GCE Source Residual}
\end{sideways}
\includegraphics[width=1.68truein]{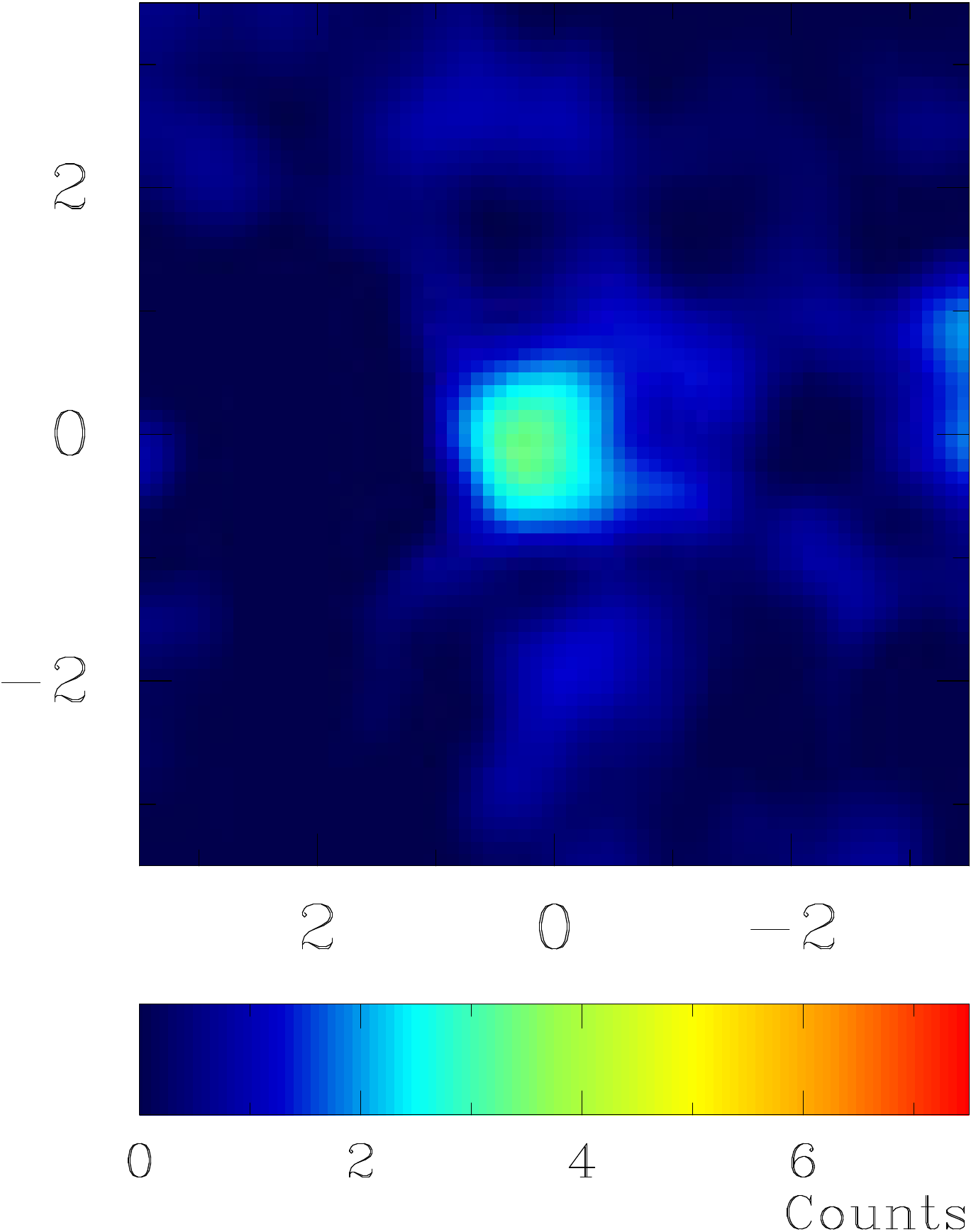}
\includegraphics[width=1.68truein]{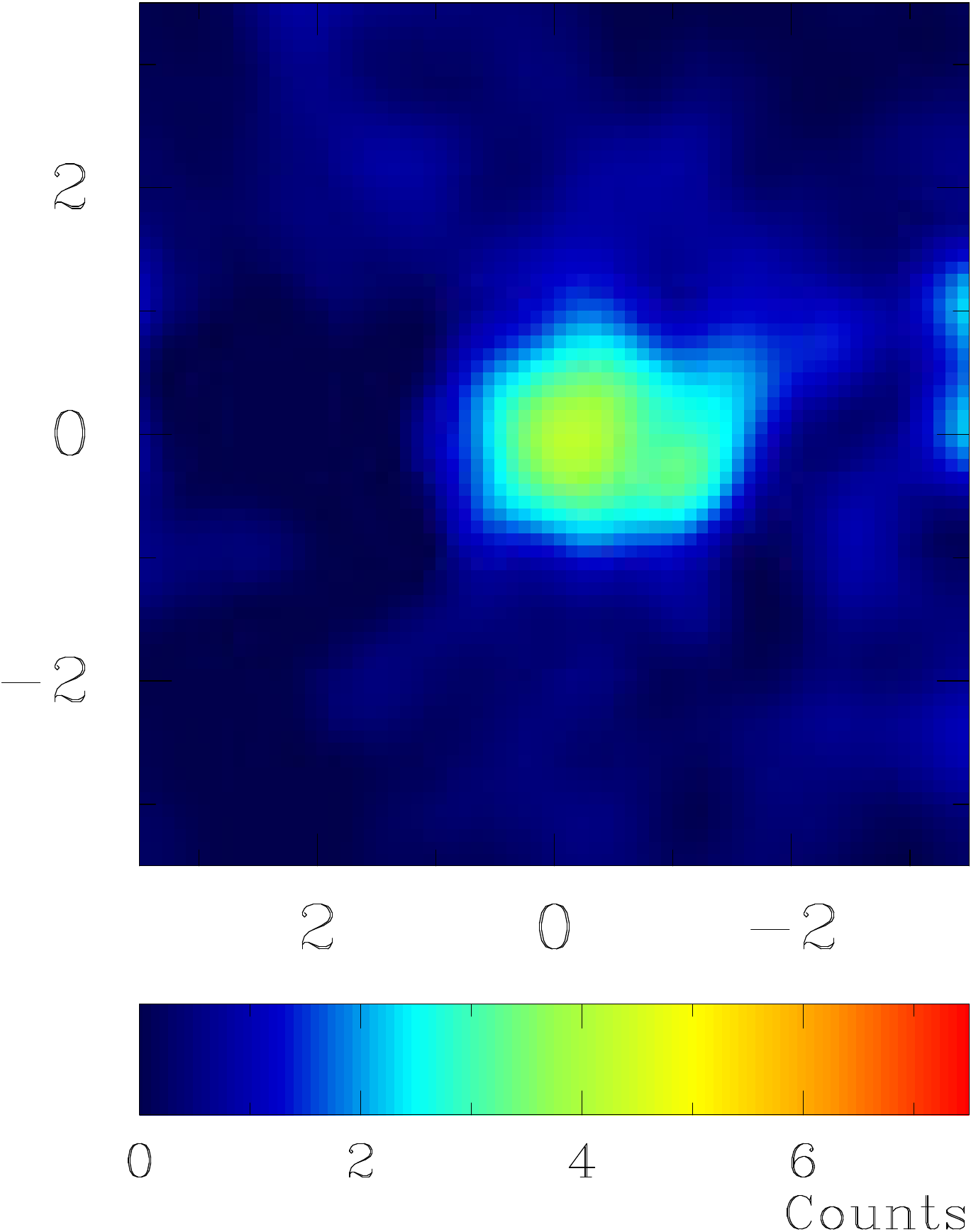}
\includegraphics[width=1.68truein]{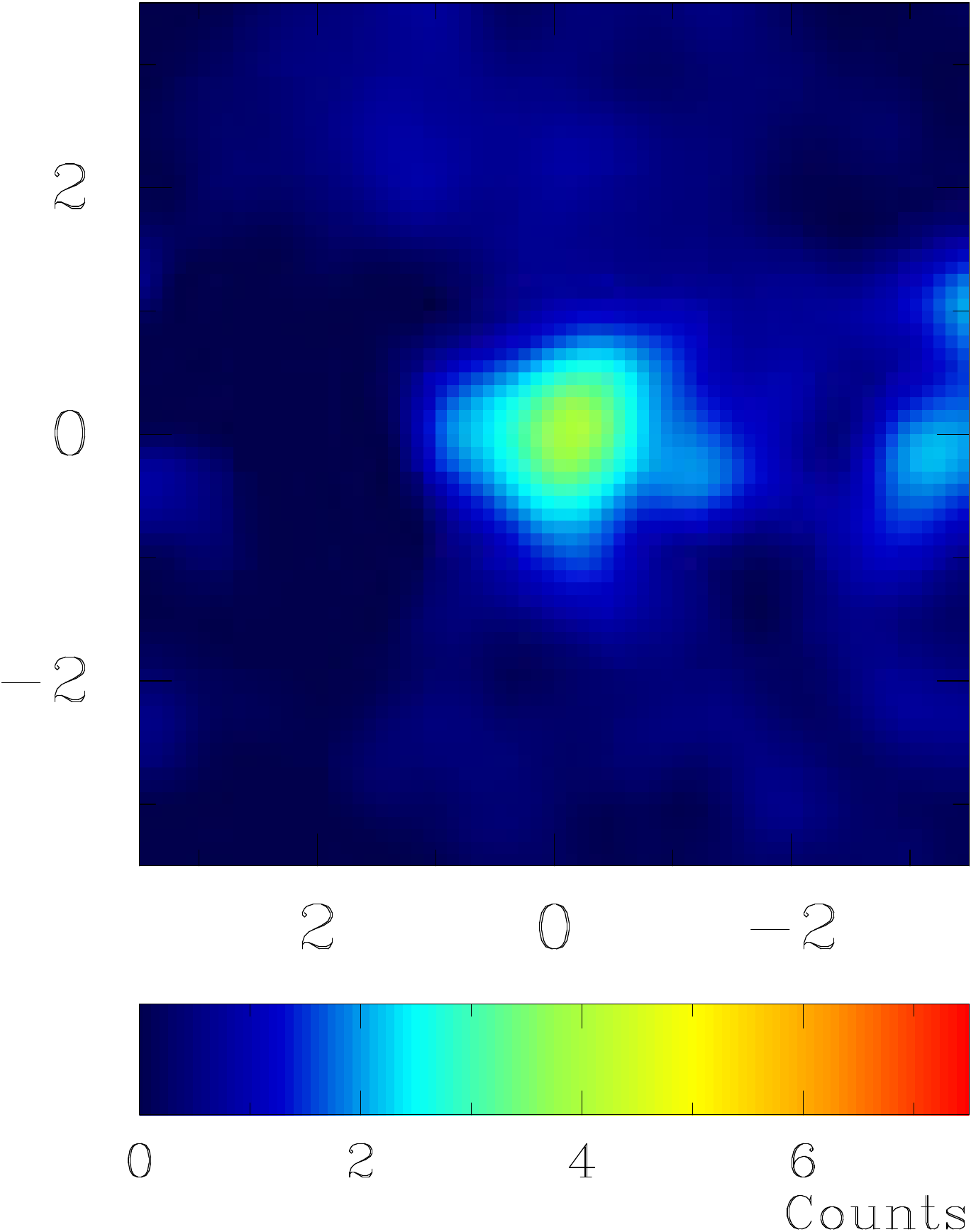}
\includegraphics[width=1.68truein]{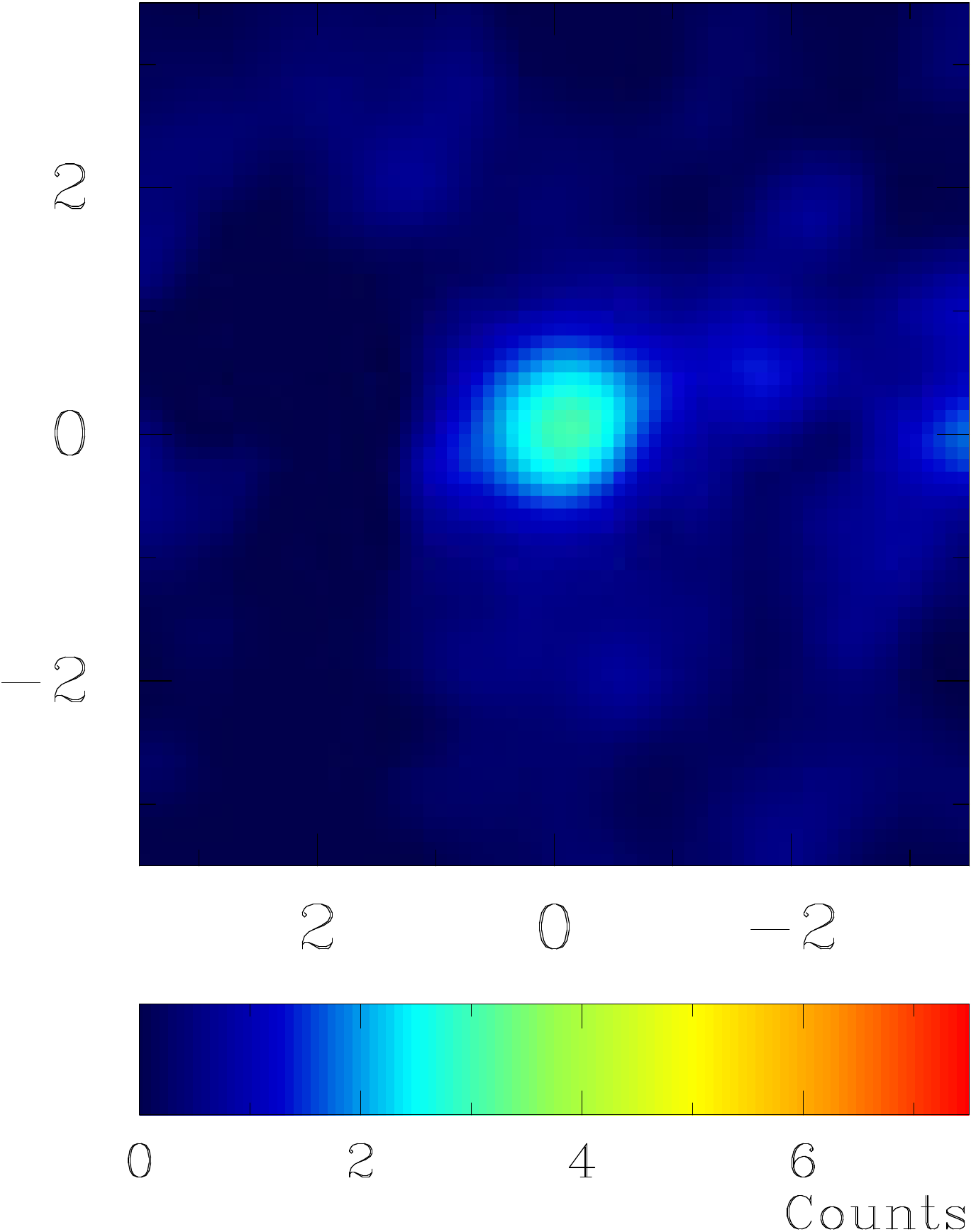}

\caption{For the full model, 2FGL+2PS+I+MG+ND+GCE (see text and Table
  \ref{logliketable}), we show here the multicomponent diffuse model
  (the combined I+MG+ND) residuals, i.e., the counts subtracting
  all model components other than the I+MG+ND components (top row),
  the multicomponent diffuse model, I+MG+ND, (middle row), and the GCE
  source residuals within our ROI (bottom row), at $|b| < 3.5^\circ$
  (vertical axis) and $|\ell| < 3.5^\circ$ (horizontal axis). The maps
  are shown on the same color scale to show the components' relative
  strength for the counts per pixel, Gaussian filtered spatially with
  $\sigma=0.3^\circ$.
\label{diffusemap}}
\end{figure*}

Alternatively, the high density of compact objects, cosmic-ray
emission, and other astrophysical activity in the GC is also expected
to be a considerable source of gamma-ray emission. The massive GC
Central Stellar Cluster may harbor a significant millisecond pulsar
(MSP) population that can have similar gamma-ray flux and spectrum as
the observed extended source in the GC \cite{Abazajian:2010zy}.  There
is also a significant detection of gamma-ray emission associated with
molecular gas as mapped by the 20 cm radio map toward the GC
\cite{YusefZadeh:2012nh}. In Ref.~\cite{YusefZadeh:2012nh}, the 20 cm map 
had the strongest statistical detection of the diffuse source templates
studied, and is interpreted as bremsstrahlung emission of high-energy
electrons interacting with the molecular gas (MG). In addition, the
gamma-ray point source associated with Sgr A$^\ast$ is among the
brightest sources in the gamma-ray sky. Sgr A$^\ast$'s spectrum from
low- to high-energy gamma rays has been modeled to originate from
cosmic-ray protons transitioning from diffusive propagation at low
energies to rectilinear propagation at high energies
\cite{Chernyakova:2011zz,Linden:2012iv}. Interestingly, the potential
confusion between pion decay, pulsar spectra and DM annihilation was
studied well before the launch of the Fermi LAT \cite{Baltz:2006sv}.

In this paper, we perform a detailed analysis of the nature of the
extended gamma-ray source from the GC region, which we designate as
the GC extended (GCE) source, the point sources in the GC, as well as
the diffuse emission associated with the 20 cm MG map. We focus on a
region of interest (ROI) of $7^\circ\times 7^\circ$ centered at the
GC. Since there have been detections of all of these sources
independently and their spatial information overlaps, we perform a new
analysis which consistently includes all of these sources---extended,
point-like, and diffuse---as well as their uncertainties determined by
the data.  In addition, including systematic and statistical
uncertainties, we determine the best fit particle masses and
annihilation channels if the GCE is interpreted as DM. Conversely, in
the case of interpreting the GCE source as an MSP population, we
discuss the number of MSPs required within our ROI and we also place
strong limits on DM annihilation cross sections.

\section{Method}
We use Fermi Tools version {\tt v9r31p1} to study Fermi LAT data from
August 2008 to May 2013 (approximately 57 months of data). We use Pass
7 rather than Pass 7 Reprocessed instrument response functions since
the latter have strong caveats for use with new extended sources. We
include point sources from the 2FGL catalog~\cite{Nolan2012} in our
ROI, $7^\circ\times 7^\circ$ around the GC centered at $b=0, \ell=0$.
Our procedure is similar to those described in
Ref.~\cite{Abazajian:2012pn}: we do two separate analyses one from 200
MeV to 300 GeV and the other including only photons with energies
between 700 MeV to 7 GeV to focus in on the energy window where the
new signal is found. We will use ``E7'' to label this analysis with
photons in the 0.7 to 7 GeV range. For the $0.2 - 300\ {\rm GeV}$
analysis, we use the {\tt SOURCE}-class photons binned in an Aitoff
projection into pixels of $0.1^\circ\times 0.1^\circ$ and into 30
logarthmically spaced energy bins. {\tt SOURCE}-class events were
chosen in order to maximize the effective area while at the same time
keeping the cosmic-ray background contamination to below the
recommended rate needed to ensure little effect on the detection and
characterization of point sources and low latitude diffuse sources, as
recommended by the Fermi Collaboration analysis documentation.

We choose the high-energy upper limit for this analysis to probe
limits on massive ($\sim$1 TeV) dark matter (see
Sec.~\ref{limitssection}).  For the $0.7 - 7\ {\rm GeV}$ analysis we use
the {\tt ULTRACLEAN}-class photons binned into pixels of
$0.2^\circ\times 0.2^\circ$ and into 12 logarthmically spaced energy
bins. In this section we describe the components of our fits.

\subsection{Fit components}

\label{astrosources}

\begin{figure*}[h!t!]
\includegraphics[width=1.72truein]{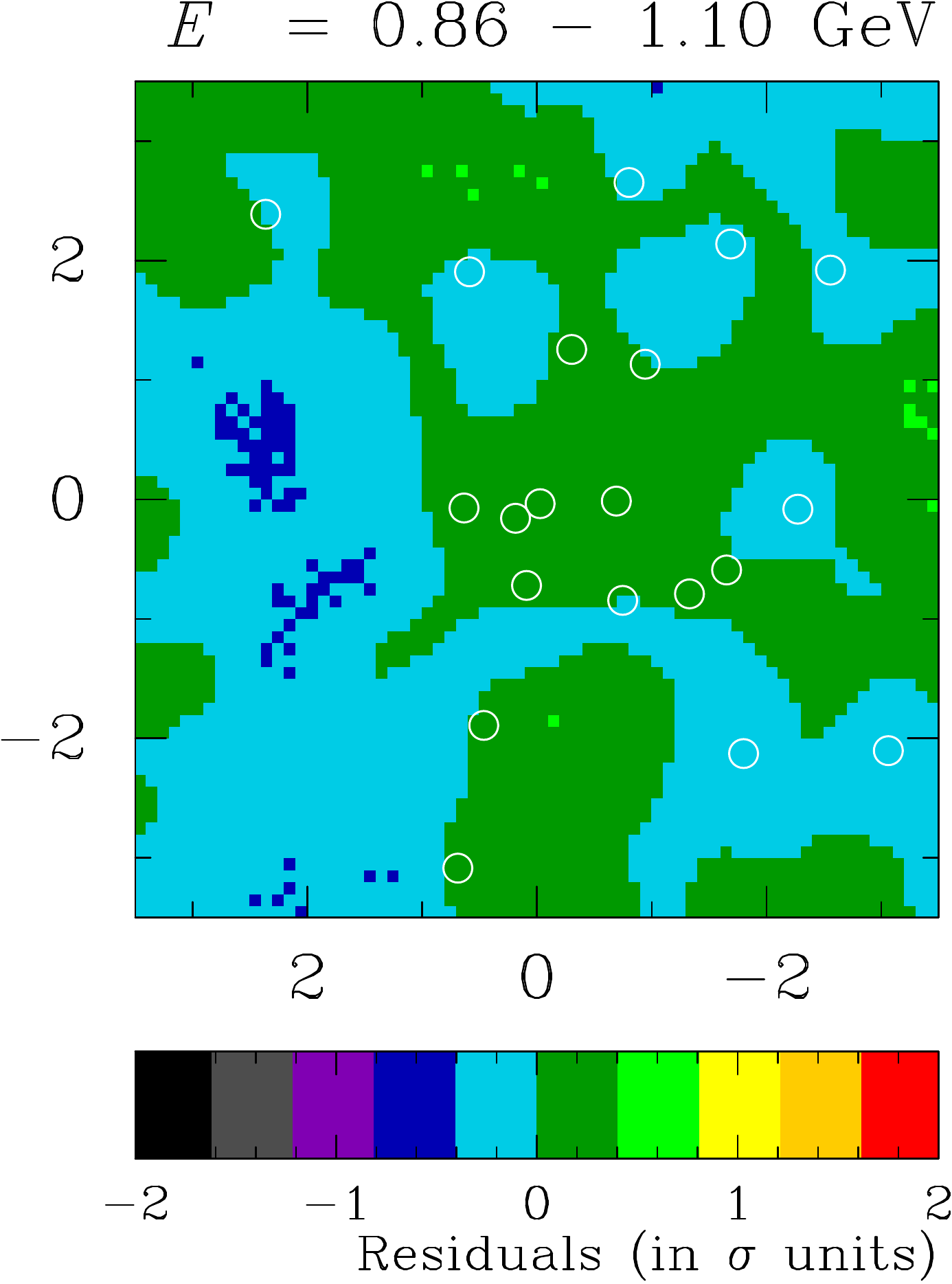}
\includegraphics[width=1.72truein]{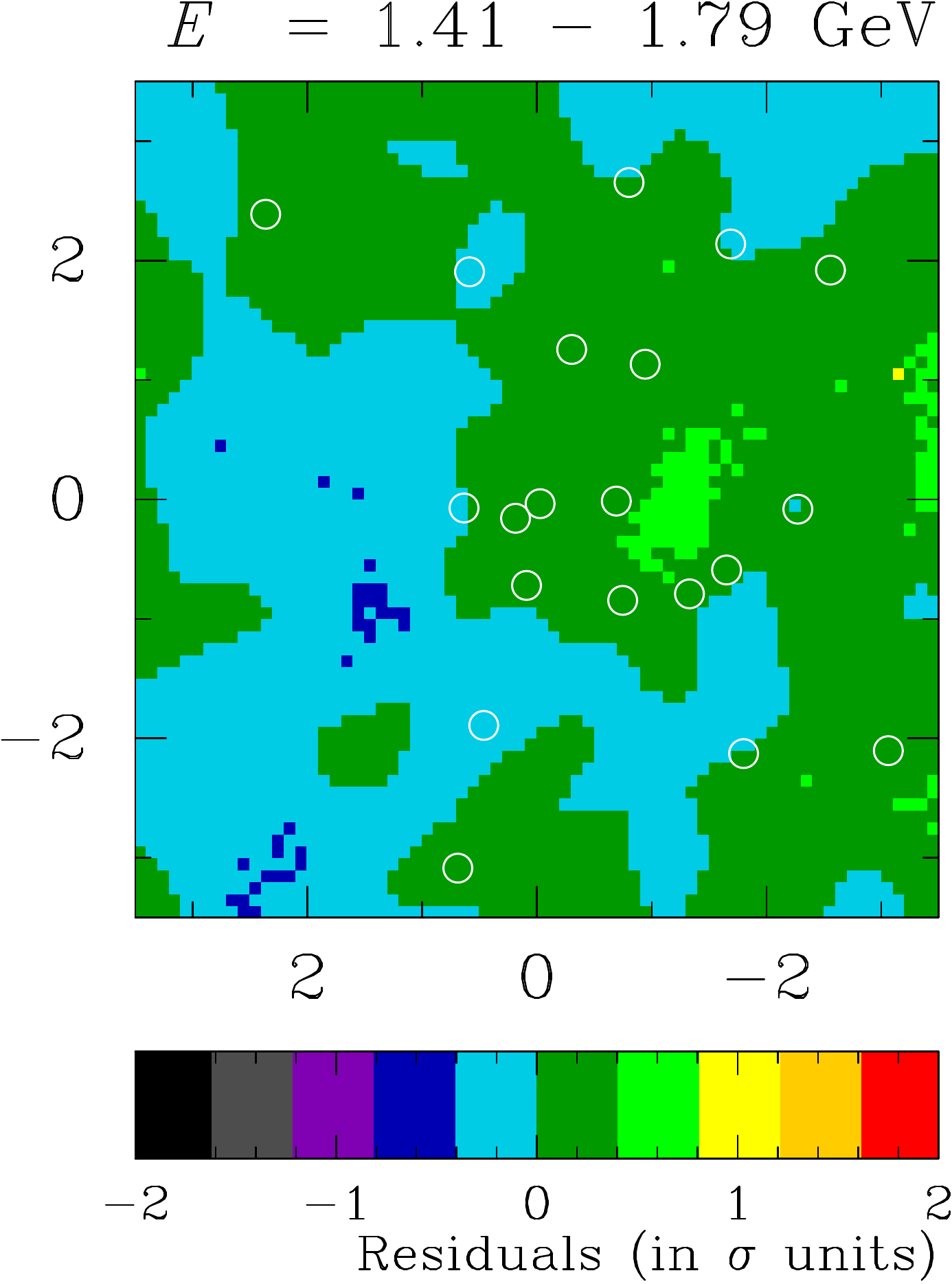}
\includegraphics[width=1.72truein]{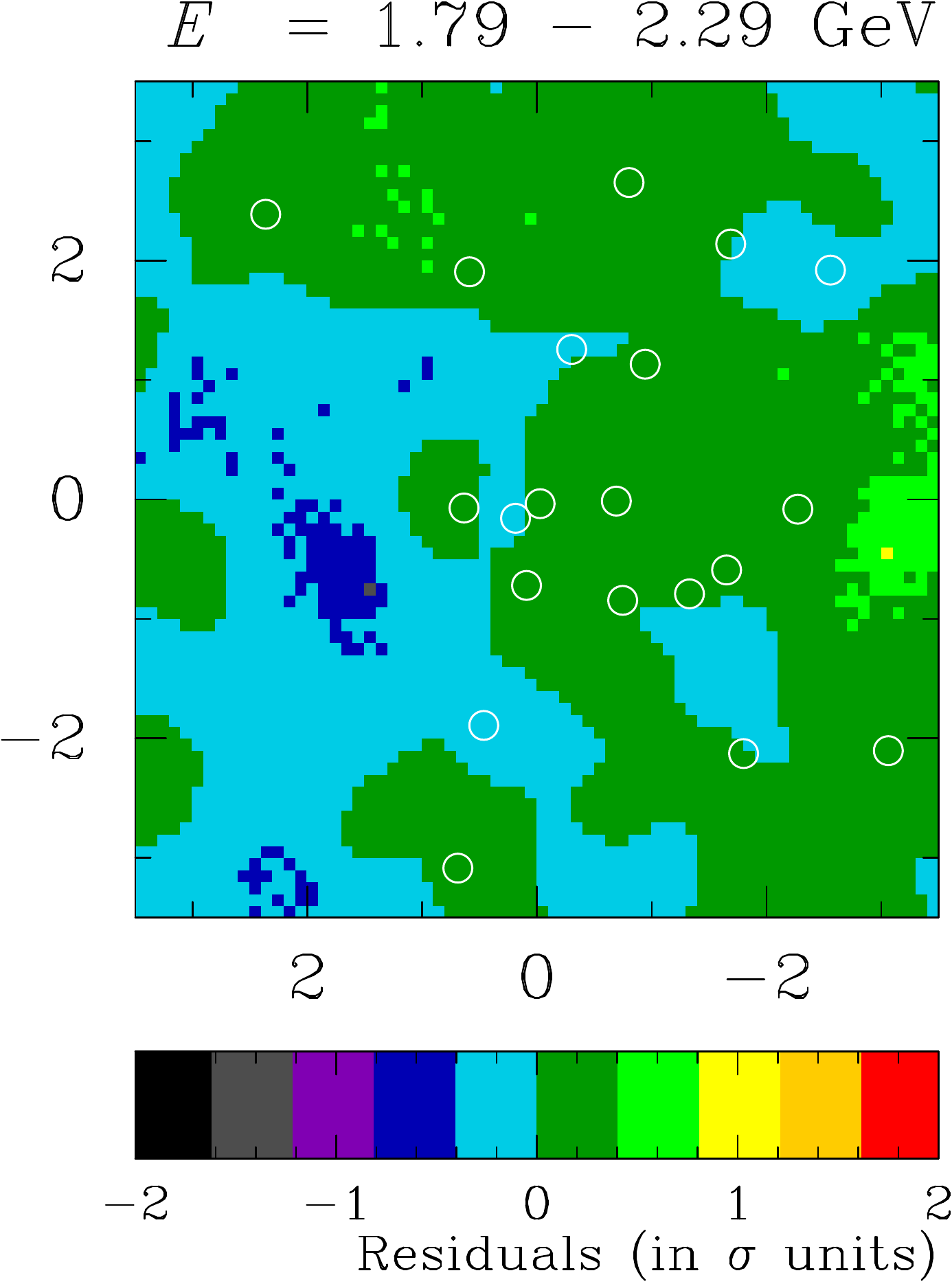}
\includegraphics[width=1.72truein]{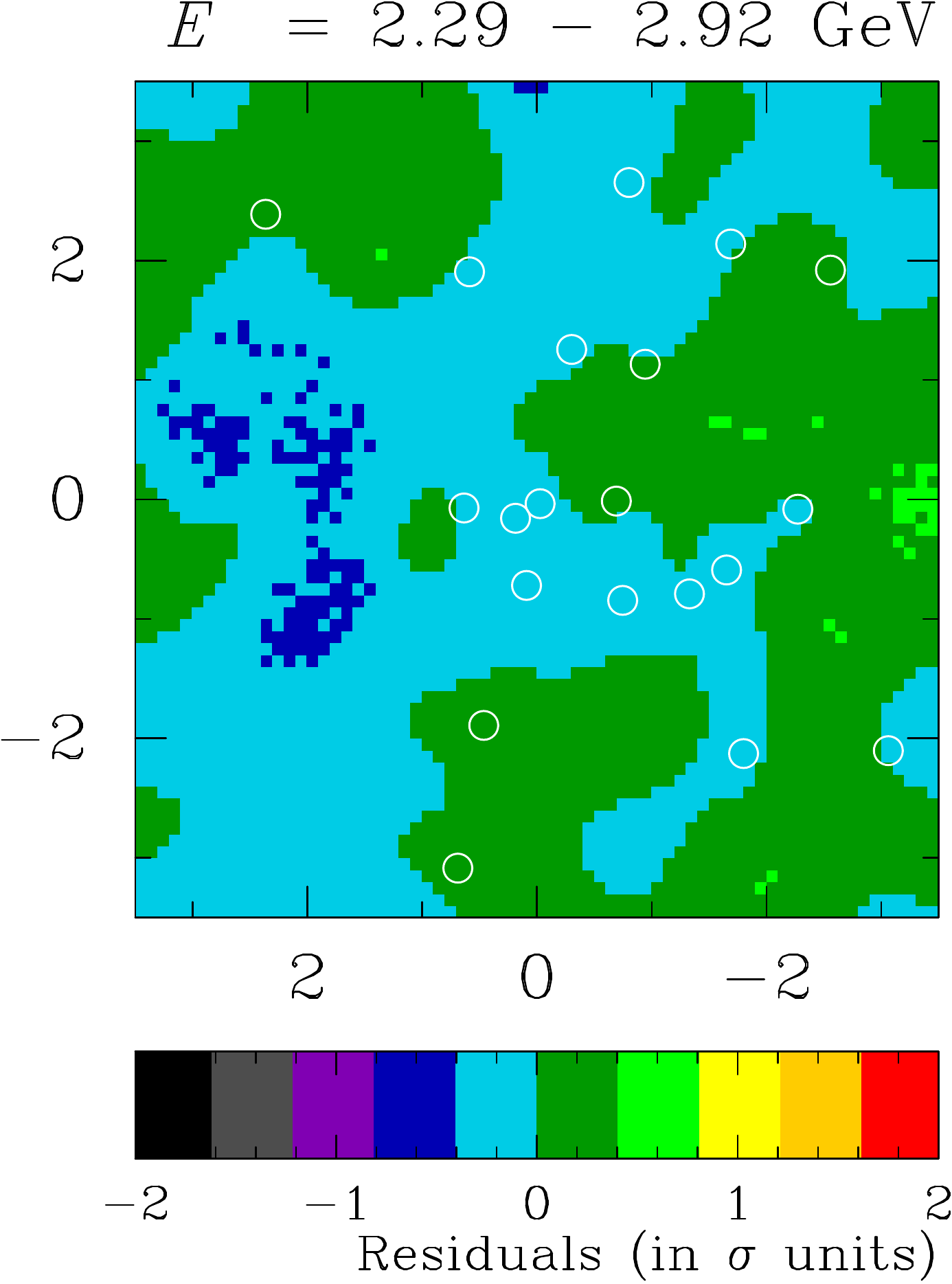}

\caption{For the full model, 2FGL+2PS+I+MG+ND+GCE (see text and Table
  \ref{logliketable}), we show the full model residuals after
  including all diffuse components, in units of $\sigma$. Here, $|b| <
  3.5^\circ$ (vertical axis) and $|\ell| < 3.5^\circ$ (horizontal
  axis). The residual count map was Gaussian filtered spatially with
  $\sigma=0.3^\circ$. The 20 point sources modeled simultaneously with
  the diffuse and extended sources in the ROI are shown as circles.
\label{residmap}}
\end{figure*}

The most minimal fitted model is based solely on the 2FGL point
sources, in addition to the recommended diffuse emission models
associated with the Galactic emission (\verb|gal_2yearp7v6_v0|) and
the isotropic background emission (\verb|iso_p7v6source|) which
includes contributions from both an extragalactic component and an
isotropic diffuse component.

Because the Galactic diffuse background is the dominant component in
the ROI, errors in the assumptions used to derive the model could
potentially have a large effect on the characterization of sources in
this region, and uncertainties associated with this component should
account for the largest source of systematic error. Here, we briefly
describe the major components that went into this model and how they
were derived. In short, the Galactic diffuse model was developed using
gas column-density maps as templates for $\pi_{0}$ decay and
bremsstrahlung emission, a model for the inverse Compton (IC) emission
calculated using GALPROP, and an intensity map for emission not traced
by the gas or IC model. These components were then fitted to
observations taken by the LAT in order to determine the emissivities
and normalization factors.  Additionally, we note that an updated
model for Pass 7 reprocessed data was released, but the Fermi
Collaboration does not recommend using this model to study gamma-ray
sources in the GC due to the inclusion of additional empirically
fitted sources at scales with extension more than 2 degrees.

Beyond the 2FGL point sources, we include two new point sources that
were detected with $\ts =2\Delta \ln(\mathcal L) > 25$, originally
found to be significant in Ref.~\cite{YusefZadeh:2012nh}. One is from
the 1FGL catalog 1FGL J1744.0-2931c, and the other is designated
``bkgA.'' We refer to the combined 2FGL and two additional point
source model as 2FGL+2PS. 

We next consider a source corresponding to emission from MG.  For its
spatial template, we use the Green Bank Telescope 20 cm radio map as
used in Ref.~\cite{YusefZadeh:2012nh}, originally from
Ref.~\cite{Law:2008uk}. The 20 cm template was originally adopted to
explain the GCE as nonthermal bremsstrahlung emission from cosmic-ray
electrons interacting with MG particles. The inclusion of the 20 cm
map is warranted due to the presence of significant features that do
not appear in the Fermi Galactic diffuse model. This is shown clearly
in Fig. 4a from Ref.~\cite{YusefZadeh:2012nh}, which shows a residual
count map after subtracting the diffuse and isotropic templates,
leaving a structure that closely traces the ridge. Consequently, the
MG template allows us to better account for the gamma-ray emission due
to high-energy processes than would be possible with the Galactic
diffuse template alone.

For the GCE source we adopt a spatial map that corresponds to a DM
density-squared template as described in Sec.~\ref{dmmodels}. As shown
below, the DM density's inner profile is dominated by a power law
increasing as $\propto r^\gamma$. When interpreted as MSP, the
real-space density corresponds to $n_{\rm MSP} \propto
\rho^{2\gamma}$.

We also test the potential presence of a diffuse (or extended) source
associated with the same density profile of the Central Stellar
Cluster of the Milky Way. To do this, we test the significance of a
source with spatial profile $n_{\rm Dif} \propto \theta^{-\Gamma}$,
where $\theta$ is the angular separation from the GC ($b=0,
\ell=0$). We designate this new diffuse source as ND below, and we
allow $\Gamma$ to vary from $-1.3$ to $+0.8$ when performing fits,
which allows for a radially decreasing (positive $\Gamma$) and increasing
(negative $\Gamma$) new diffuse component.

We find that the fitted normalization of the isotropic background
emission, \verb|iso_p7v6source|, is significantly higher than unity
for all model cases. Therefore, we perform fits with the isotropic
background emission model \verb|iso_p7v6source| fixed to unity but
with an added new isotropic component (denoted ``I'' in the model
names) over the ROI with a free power-law spectrum. The reason we fix
the isotropic background model is because it is meant to account for
extragalactic diffuse gamma rays and misclassified cosmic rays, and so
should not depend strongly on the chosen ROI. We emphasize that all other
parameters for model components within the ROI, including diffuse and
point sources, were varied during the fitting procedure.

We refer to the new isotropic diffuse model, I, together with the new
MG and the ND components, as the ``multicomponent diffuse model''.  In
the top row of Fig.~\ref{diffusemap} we show the residual for the new
diffuse models, i.e., the raw counts minus a model that includes all
components except the multicomponent diffuse model. With inclusion of
all components, no significant major residuals are found, as shown in
Fig.~\ref{residmap}. One region of negative residual is seen at
$b=-1^\circ$, $\ell=+2^\circ$ that is associated with a feature at
that position in the \verb|gal_2yearp7v6_v0| Galactic diffuse model.

The combination 2FGL+2PS+I+MG+ND+GCE defines our \emph{full model}
(bottom row of Table \ref{logliketable}).  When fitted, the new
isotropic diffuse component (I) is found with high statistical
significance with a flux that is 1.4 times that of the two-year Fermi
isotropic background model \verb|iso_p7v6source| within our ROI.  The
spectrum is similar to that of \verb|iso_p7v6source| with a power law
index of $1.980\pm 0.082$. For the E7 (0.7 to 7 GeV) analysis we did
not include a new power-law isotropic source but instead let the
normalization of \verb|iso_p7v6clean| vary since the two are so
similar to each other.

In addition to these sources, we also ran the Fermi tool {\tt gttsmap}
with a coarse binning of $0.4^\circ$. Given the high counts with the
ROI we expected to pick up a lot of structure so we restricted our
search to within the inner $4^\circ\times4^\circ$.  The map of TS
values does indeed have many pixels with TS $> 25$ but most of them
are likely not point sources. We picked the pixels with TS $>45$ and
added them as point sources to the E7-2FGL+2PS+MG+GCE. The fit
constrained six of these putative point sources and the total fit
improved by $\Delta \ln \mathcal L = 110$ due to the addition of these point
sources. We urge caution in interpreting these six new sources as bona
fide point sources since that requires a more detailed analysis with
finer binning. Our main aim here is to construct an empirical model of
the emission and adding these sources definitely helps. We have not
added these sources to the $>200$ MeV analysis since they were found
in the more restricted energy window. There were also no significant
changes to the GCE spectrum with the addition of these sources. We
will refer to these sources (added as point sources) as nPS.

Since the GC region is bright, we have redone the analysis and
modeling using only Fermi LAT front-converting photons
(\verb|P7SOURCE_V6::FRONT|), and find very similar results to the full
data set. The TS of the GCE source goes from 170.7 for the full data
to 156.7 with \verb|FRONT| converting data alone, and the other diffuse
and point sources are not significantly affected. The normalization
and spectrum of the GCE source does change, with the full data set
giving the GCE a flux of $(3.1\pm 0.3)\times
10^{-7}\rm\ ph\ cm^{-2}\ s^{-1}$ and log-parabola parameters of
$\alpha = -4.28 \pm 0.18$, and $\beta = 0.959 \pm 0.026$, while the
\verb|FRONT| data set gives the softer spectrum $\alpha = -1.15 \pm
0.10$, and $\beta = 0.507 \pm 0.017$ with a higher flux of $(7.1\pm
0.8)\times 10^{-7}\rm\ ph\ cm^{-2}\ s^{-1}$, mostly attributable to
more low-energy photons in the softer spectrum. We show the
\verb|FRONT| converting photon residual GCE spectrum in
Fig.~\ref{GCEspectrum}. The systematic shift for the \verb|FRONT|
analysis is indicative of the systematic uncertainty in determining
the GCE spectrum which is strongly degenerate with the other diffuse
and point sources, and which also depends on the assumed spectrum
(Fig.~\ref{systematicsspectrum}) and the nature of the MG model
(Fig. \ref{GCEspectrum}).

\begin{table}[t!]
\caption{Models' renormalized log likelihood values, as reported by
  the Fermi Science Tools, $-\ln [\mathcal L\times(\sum_i k_i!)) ]$,
  where $k_i$ is the photon count in bin $i$, for the various models
  and the $\Delta \ln(\mathcal L)$ as compared to the 2FGL-only model for
  the analysis where photons in the energy range 0.2 to 300 GeV were 
  included. The model in the last row, 2FGL+2PS+I+MG+ND+GCE, defines our 
  full model.}
\label{logliketable}
\begin{ruledtabular}
\begin{tabular}{l|cc}
 Model & $-\ln[\mathcal L\!\!\times\!\!(\sum_i k_i!) ]$ & $\Delta \ln{\mathcal L}$ \\
 \hline \\
 {\small 2FGL\footnotemark[1]} & -1080408.3 & -- \\
 {\small 2FGL+2PS\footnotemark[2]} & -1080510.3 & 102.0 \\
 {\small 2FGL+2PS+I\footnotemark[3]}	& -1080685.7 & 277.4 \\
 {\small 2FGL+2PS+I+MG\footnotemark[4]} & -1080931.1 & 522.8 \\
 {\small 2FGL+2PS+I+MG+ND\footnotemark[5]}  $\Gamma\!=\!-0.5$& -1081012.9 & 604.7 \\
 {\small 2FGL+2PS+I+MG+GCE\footnotemark[6]} $\gamma\!=\!1.1$ & -1081061.5 & 653.2 \\
 {\small 2FGL+2PS+I+MG+GCE $\gamma\!=\!1.1$}\\{\small $\qquad$ + ND $\Gamma\!=\!-0.5$ }& -1081098.3 & 690.0 \\
 \hline
\end{tabular}
\end{ruledtabular}
\footnotetext[1]{Point sources in the 2FGL catalog, together with \texttt{gal\_2yearp7v6\_v0} and \texttt{iso\_p7v6source} diffuse models}
\footnotetext[2]{The two additional point sources (PS) found in the ROI}
\footnotetext[3]{The new isotropic component (I) with free power-law spectrum; note \texttt{iso\_p7v6source} is kept fixed when this is added. }
\footnotetext[4]{The 20 cm radio map template (MG)}
\footnotetext[5]{The new diffuse model (ND) with its respective $\Gamma$}
\footnotetext[6]{The Galactic Center excess (GCE) with its respective $\gamma$}
\end{table}

Several point sources as well as the diffuse and extended sources
associated with the MG and GCE source emission are fit with
``log-parabola'' spectra of the form
\begin{equation}
\frac{dN}{dE} =
N_0\ \left(\frac{E}{E_b}\right)^{-(\alpha+\beta\ln(E/E_b))},
\end{equation}
keeping $E_b$ fixed, yet source dependent, and fitting the other
parameters $\alpha$, $\beta$, and $N_0$.

\subsection{Dark matter models}
\label{dmmodels}

\begin{figure}[t]
\begin{center}
\includegraphics[width=3.4truein]{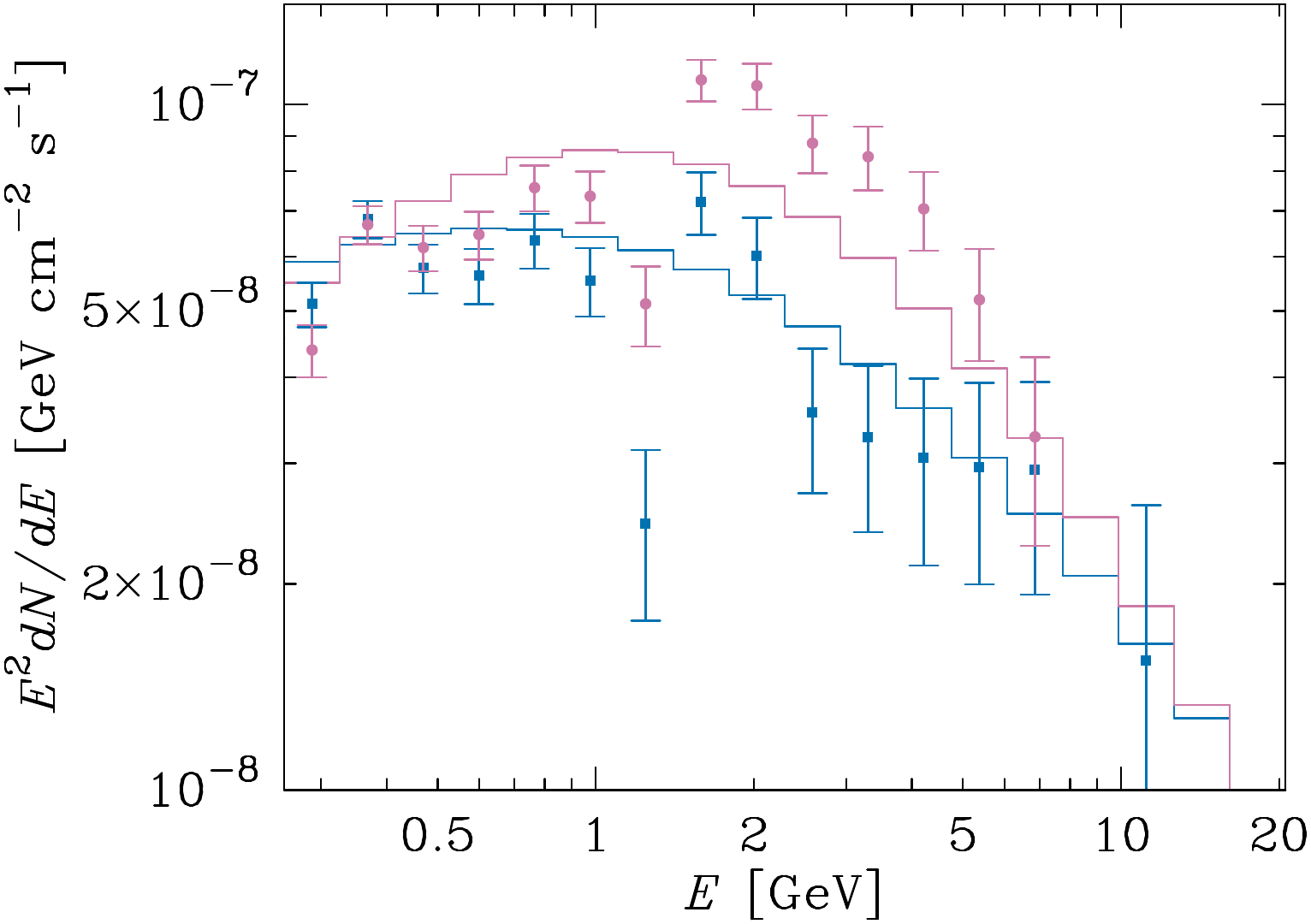}
\end{center}
\caption{Shown are two cases of our determination of the Sgr A$^\ast$
  source spectrum. The 2FGL+2PS+I binned spectrum is in pink circles,
  with best fit binned log-parabola spectrum in pink. The full model
  2FGL+2PS+I+MG+ND+GCE spectrum is in blue squares, with best fit binned
  log-parabola spectrum in blue. The presence of GCE associated
  photons at 1 to 3 GeV in the Sgr A$^\ast$ spectrum is evident in the
  case of the 2FGL+2PS+I modeling. The errors shown are solely the
  Poisson errors within the energy band and do not reflect covariances
  or systematic uncertainties.  \label{SgrAresidspectrum}}
\end{figure}

\begin{figure}[t]
\begin{center}
\includegraphics[width=3.4truein]{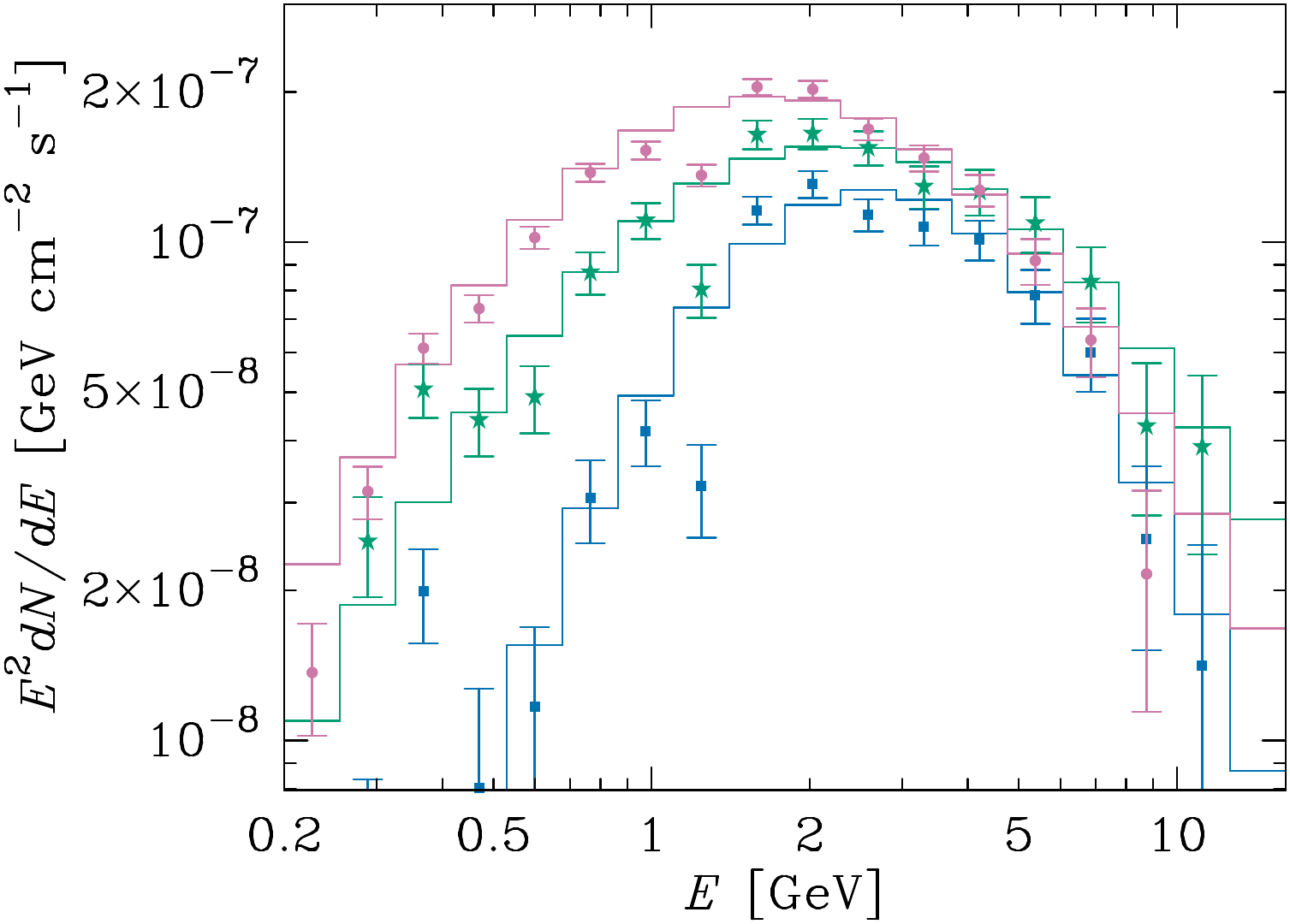}
\end{center}
\caption{Shown are two cases of our determination of the GCE source
  spectrum. The 2FGL+2PS+I+GCE binned spectrum is in pink circles, with best
  fit binned log-parabola spectrum in pink. The full model
  2FGL+2PS+I+MG+ND+GCE spectrum is in blue squares, with best fit binned
  log-parabola spectrum also in blue. We also show the spectrum using
  {\tt FRONT} converting only photons in green stars, with its best
  fit binned log-parabola spectrum in green. The errors shown are solely the
  Poisson errors within the energy band and do not reflect covariances
  or systematic uncertainties.  \label{GCEspectrum}}
\end{figure}

For the GCE source, we employ spatial templates derived from
``$\alpha\beta\gamma$'' profiles fashioned after the
Navarro-Frenk-White (NFW) profiles
\cite{Navarro:1996gj,Klypin:2001xu},
\begin{equation} \label{nfw}
\rho\left(r \right) =
\frac{\rho_\text{s}}{\left(r/r_\text{s}\right)^\gamma
  \left(1+\left(r/r_\text{s}\right)^\alpha\right)^{(\beta-\gamma)/\alpha}}
\end{equation}
with fixed halo parameters $\alpha=1$, $\beta=3$, $r_s =
23.1\rm\ kpc$, and a varied $\gamma$ inner profile.  The canonical NFW
profile has $\gamma\equiv 1$. Note, the parameters $\alpha$ and $\beta$ here
are never varied.

The differential flux for a dark matter candidate with cross section
$\langle\sigma_{\rm A}v\rangle$ toward Galactic coordinates $(b,\ell)$
is
\begin{equation}
\frac{d\Phi(b,\ell)}{dE}=\frac{\langle\sigma_{\rm A}v\rangle}{2}\frac{J(b,\ell)}{J_0}\frac{1}{4\pi m_\chi^2}\frac{dN_\gamma}{dE}\enspace,
\end{equation}
where $dN_\gamma/dE$ is the gamma-ray spectrum per annihilation and
$m_\chi$ is the dark matter particle mass.  The quantity $J$ is the
integrated mass density squared along line of sight, $x$,
\begin{equation}
J(b,\ell)=J_0 \int d\,x\ \rho^2(r_{\rm gal}(b,\ell,x))\enspace,
\end{equation}
where distance from the GC is given by
\begin{equation}
r_{\rm gal}(b,\ell,x)=\sqrt{R_{\odot}^2-2 x
  R_{\odot}\cos(\ell)\cos(b)+x^2}\enspace.
\end{equation}
Here, $J_0 \equiv 1/[8.5\ \rm kpc (0.3\ GeV\ cm^{-3})^2]$ is a
normalization that makes $J$ unitless and cancels in final expressions
for observables.  The value for the solar distance is taken to be
$R_\odot = 8.25\rm\ kpc$~\cite{Catena:2009mf}.  The density $\rho_s$
for the $\alpha\beta\gamma$ profile is a normalization
constant determined uniquely by the local dark matter density, $\rho_\odot$.

\subsection{Method}

In order to find the best fit models, and quantify the systematic
error inherent in the model-choice dependence in the analyses, we
found fits to a very large number of diffuse and extended source model
combinations.  Our 2FGL+2PS+I model consists of all
the 2FGL sources plus the two additional point sources, 1FGL
J1744.0-2931c and bkgA, and the new isotropic component.  We add
to this the MG template and the GCE template individually and then
together to test the significance of their detection. Then, we include
the ND model and simultaneously vary the density squared $\gamma$ and
2D projected $\Gamma$ to find the best fit morphologies for these sources.

For each of the model combination cases, we scan the dark matter
particle mass for WIMPs annihilating into $b \overline{b}$, $\tau^{+}
\tau^{-}$, and a mixture of both channels to find the best fit
particle masses. To do this, we add to each model a dark matter source
with a $\rho^{2}$ spatial template, Eq.~\eqref{nfw}, and spectrum
generated via {\sc PYTHIA} as in
Refs.~\cite{Abazajian:2010sq,Abazajian:2011ak}. For finer mass
binning, we use gamma-ray spectra generated with DarkSUSY
\cite{Gondolo:2004sc} and micrOmegas~\cite{Belanger:2013oya}. Due to
the finite intervals between particle masses, we determine the best
fit masses and errors for the various mass cases with a fourth order
spline interpolation. As can be seen in Fig.~\ref{massfit}, this 
method is sufficiently accurate. For each particle mass, we vary all
of the model parameters for the Galactic diffuse model, all new added
diffuse sources, and all point sources with TS $> 25$. We repeat this
procedure for several different models: for 2FGL+2PS+I+GCE (only point
sources and diffuse backgrounds), 2FGL+2PS+I+MG+GCE (with the MG
template included), and 2FGL+2PS+I+MG+ND+GCE (the full model, adding
both the MG and new diffuse components).

Note that the prompt spectrum produced by the particle annihilation
into both $b$ quarks and $\tau$ leptons can be significantly modified
by bremsstrahlung of the annihilation cascade particles on the dense
gas in the GC region~\cite{Cirelli:2013mqa}. The precise nature and
magnitude of the bremsstrahlung modification of the gamma-ray spectra
have a high astrophysical model dependence. In Sec.~\ref{dmsection}
below, we describe a test of the bremsstrahlung effects on the observed 
spectra and their impact on our results.

To illustrate the nature of the sources nearest the GC, we calculate
the spectrum of the source associated with Sgr A$^\ast$. We compute
the spectra by creating residual maps for the point source or extended
source of interest summing the pixel-based flux (counts divided by
exposure) in each energy bin in the residual map of the particular
source, using the inner $3^\circ\times 3^\circ$ of the ROI in order to
exclude residuals in the outer regions of the ROI. The spectrum for
Sgr A$^\ast$ and the GCE source are shown in
Figs.~\ref{SgrAresidspectrum} and \ref{GCEspectrum}.

\section{Results \& Discussion}
\label{results}
Due to the high density of sources---point, extended, and diffuse 
backgrounds---in the GC region, the inferred nature of cataloged 
point sources, new point sources, and extended sources depend
significantly on the assumed point, extended, and diffuse models. 
Below, we focus on implications for astrophysical sources, and on
the GCE source as interpreted as DM annihilation.

\subsection{Diffuse sources and Sgr A*}
\label{astrosection}

\begin{figure}[t]
\begin{center}
\includegraphics[width=3.4truein]{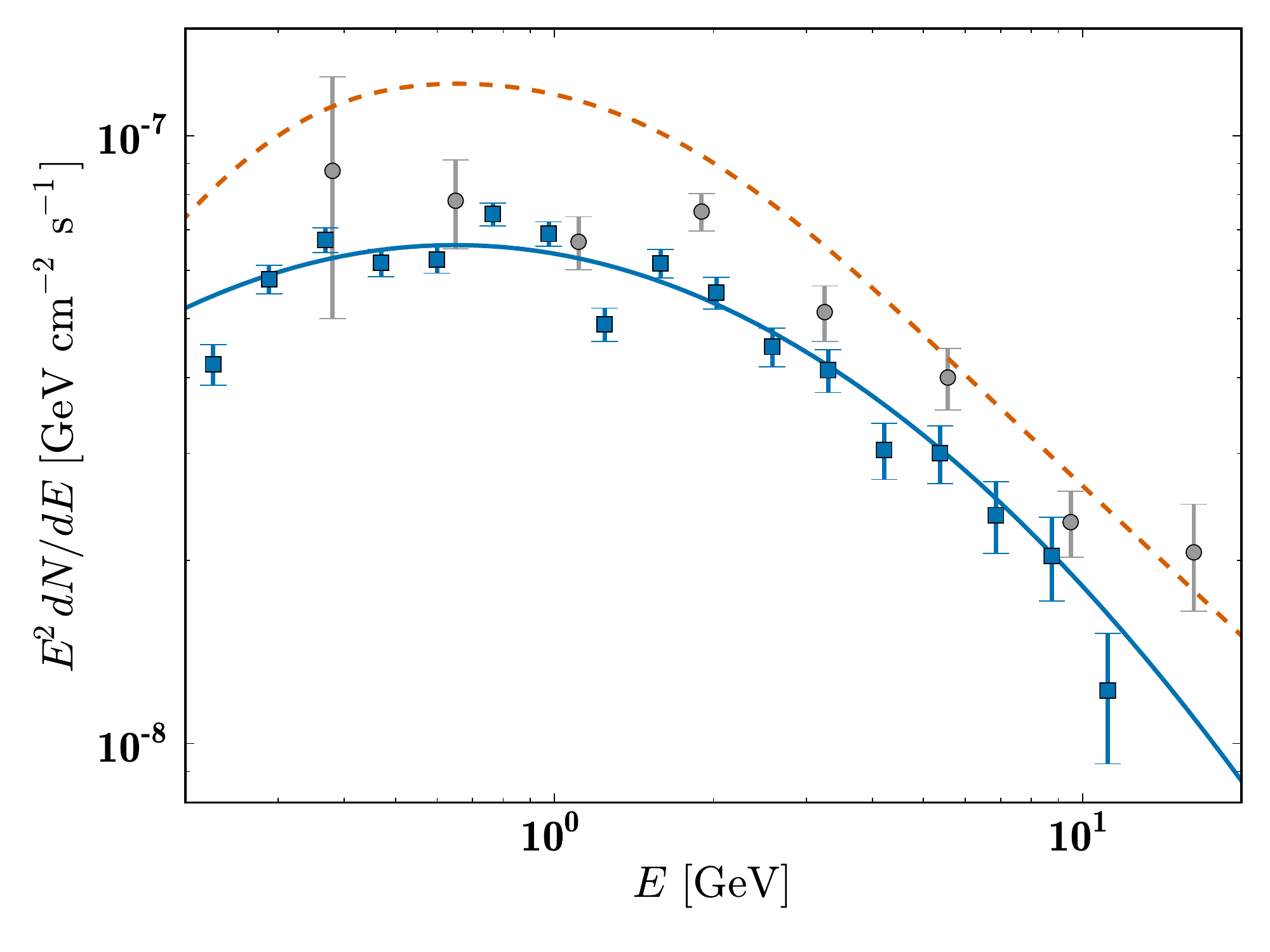}
\end{center}
\caption{Here we show the SED of the Sgr A$^\ast$ source for the full model, 
  2FGL+2PS+I+MG+ND+GCE (blue squares), as well as its best
  fit log-parabola spectrum (solid line). For comparison, we show the Sgr
  A$^\ast$ spectrum determined by Chernyakova {\it et
  al.}~\cite{Chernyakova:2011zz} (gray circles) and the 3 pc diffusion
  emission model from Linden {\it et al.}~\cite{Linden:2012iv} (dashed
  line).  The errors represent the SED-normalization statistical
  uncertainty within an energy band. \label{SgrAspectrum}}
\end{figure}

\begin{figure}[t]
\begin{center}
\includegraphics[width=3.4truein]{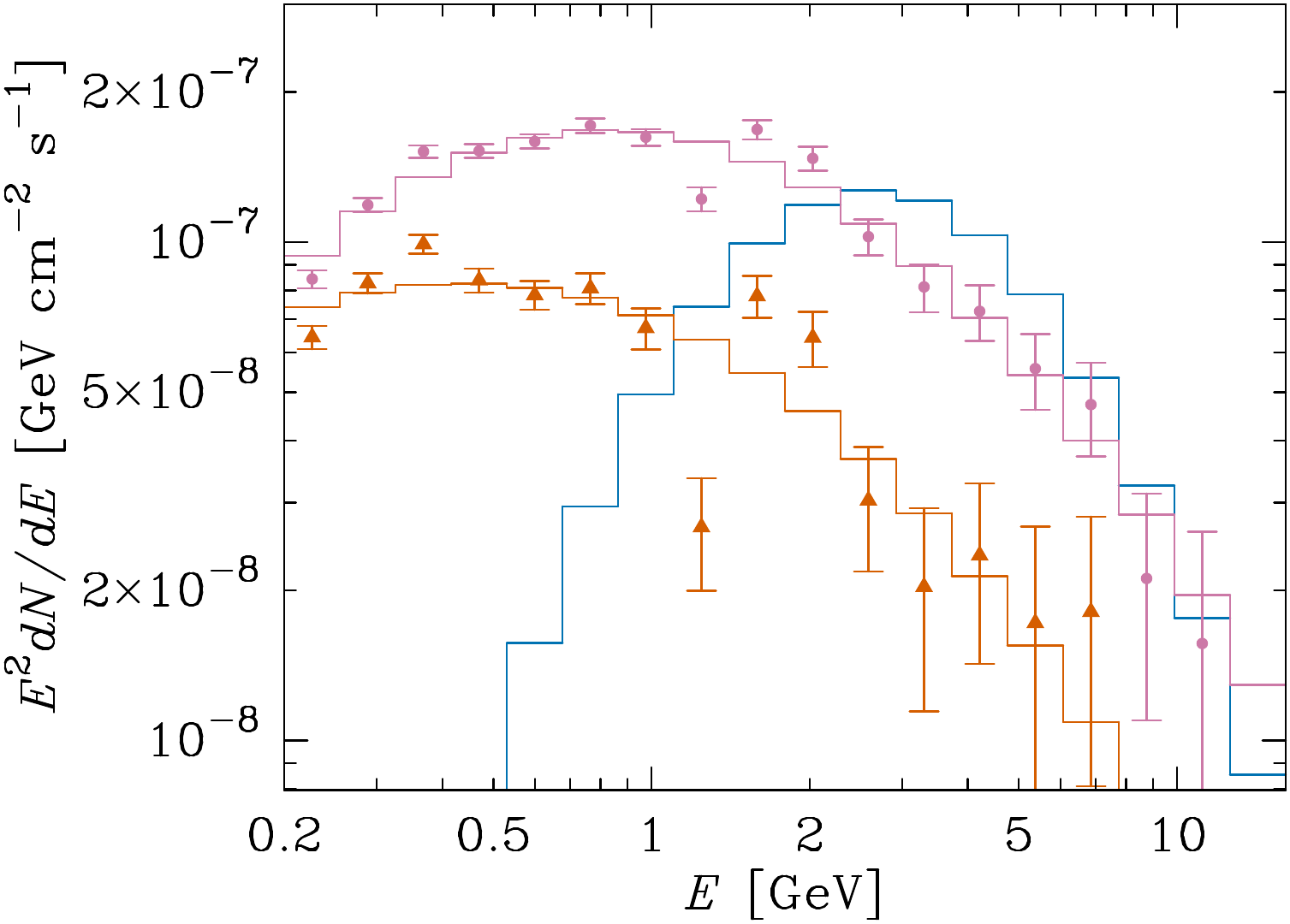}
\end{center}
\caption{Here we show the spectrum for the MG and ND components for
  the 2FGL+2PS+I+MG+ND+GCE model. The MG spectrum is in pink circles,
  with the best fit log-parabola spectrum in pink. The ND spectrum is in
  orange triangles, with the best fit log-parabola spectrum in orange. For
  reference, we show the best fit GCE spectrum for the same full
  model, which shows how the GCE is detected at above $\sim$2 GeV. The
  errors shown are solely the Poisson errors within the energy band
  and do not reflect covariances or systematic
  uncertainties. \label{MG+ND_residspectrum}}
\end{figure}

\begin{figure}[t]
\begin{center}
\includegraphics[width=3.4truein]{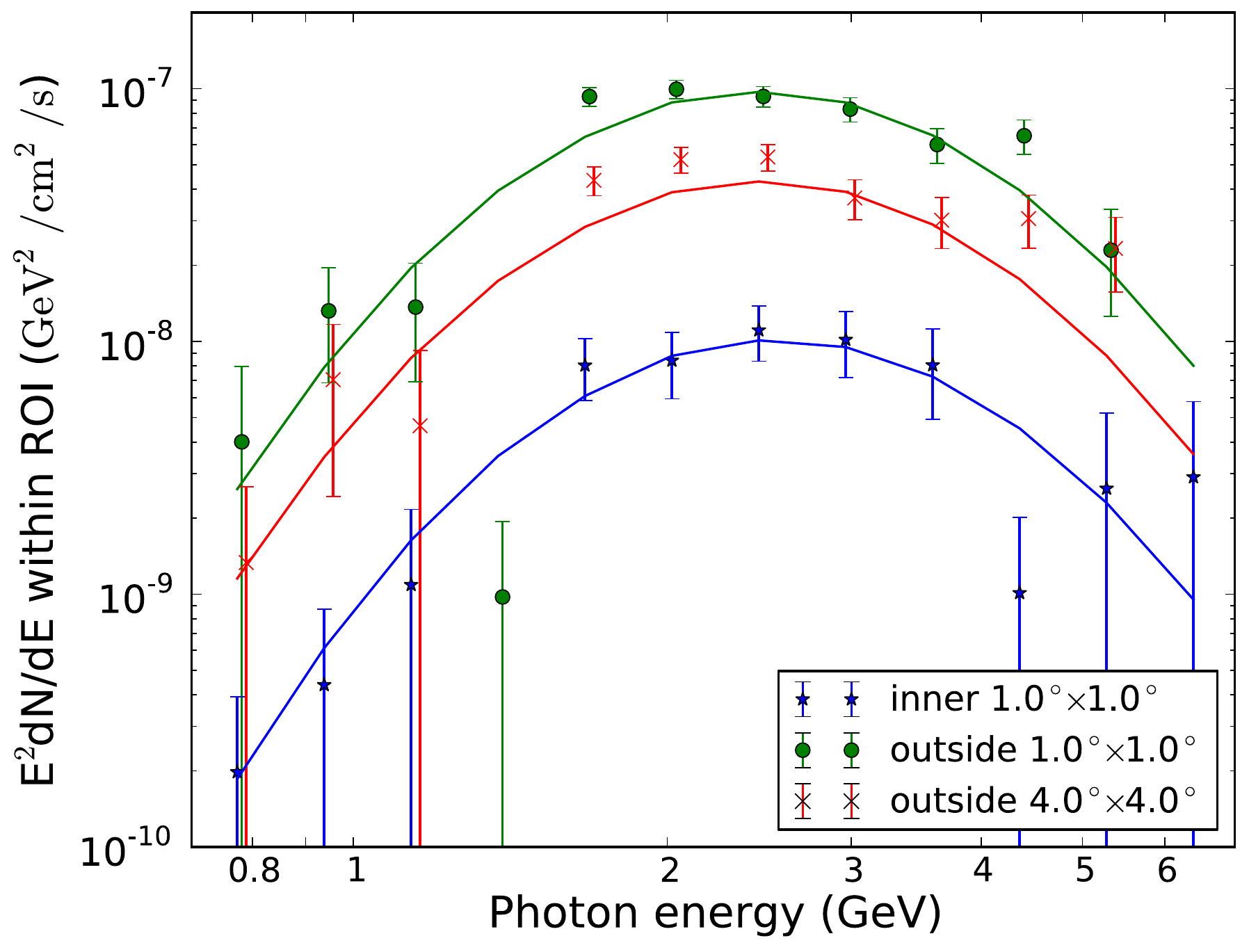}
\end{center}
\caption{Here we show the residual flux for the GCE for different spatial 
regions within the ROI for the 2FGL+2PS+MG+GCE model as well as 
the flux from the model counts for $\gamma=1.1$. It is clear that all the 
different regions are being well fit by the NFW-like density profile. 
The errors shown are solely the Poisson errors within the energy band 
and do not reflect covariances or systematic uncertainties. \label{GCEprofile}}
\vskip -0.3cm
\end{figure}

We included a number of new diffuse and extended sources in this
analysis, which were detected at high significance. First, the 20 cm
MG map was included.  The MG component was detected at a $\ts$ of
245.4 relative to the model with just the 2FGL+2PS+I sources. Second,
we added a $\rho^{2}$ GCE template and a two-dimensional projected
density profile (ND) and then scanned the morphological parameter
space of these components in $\gamma$ and $\Gamma$ for each case
separately and in combination, with $\Delta\gamma$ and $\Delta\Gamma$
scan step sizes of 0.1, leading to over four dozen morphological model
tests. The likelihood is shallow in $\Delta\Gamma$ near its minimum:
$\Delta\ln{\mathcal L}\approx 0.2$ for $\Delta\Gamma=\pm 0.1$ from
their best fit values. The change for $\Delta\gamma = \pm 0.1$ is
larger. Fitting a polynomial to the profile likelihood on the
variation of $\gamma$, we find $\gamma = 1.12\pm 0.05$ (statistical
errors only).

When both the ND and GCE sources are included, i.e.,
2FGL+2PS+I+MG+ND+GCE, and their respective indices varied, we found
that the best fit values were for $\gamma=1.1$ and $\Gamma=-0.5$,
which resulted in a $2\Delta \ln(\mathcal L)$ of 334.4 over the model
that included neither source, which indicates a strong preference for
both of these components in combination. Note that the negative
$\Gamma$ indicates a radially increasing new diffuse (ND)
component. Table~\ref{logliketable} shows the $\ln(\mathcal L)$ for
the various models as well as the $\Delta \ln(\mathcal L)$ as compared
to the 2FGL only model. Table~\ref{fluxtable} shows the flux and $\ts$
for the main extended sources and four point sources nearest to the
Galactic Center.

Including the ND source without the MG or GCE sources is a
significantly poorer fit overall since it is not as centrally
concentrated as the MG and GCE templates. Therefore, we do not
consider this model case further.

Very significantly, the presence of the GCE, MG, and ND diffuse
sources affects the inferred properties of the central point sources,
particularly Sgr A$^\ast$, as shown in
Fig.~\ref{SgrAresidspectrum}. In the 2FGL+2PS model, Sgr A$^\ast$ has
a total flux of $(3.13\pm 0.16)\times
10^{-7}\rm\ ph\ cm^{-2}\ s^{-1}$, and a curved spectrum that is
consistent with the features seen in previous work by Chernyakova {\it
  et al.}~\cite{Chernyakova:2011zz}, with a log-parabola spectrum of
$N_0= (3.112\pm 0.068)\times 10^{-11}\rm\ MeV^{-1}\ cm^{-2}\ s^{-1}$,
$\alpha = 2.242\pm 0.025$, $\beta = 0.273\pm 0.018$. However, with the
inclusion of the detected GCE source as well as MG and ND sources, Sgr
A$^\ast$ is less peaked.  The GCE shows a peaked spectrum
(Fig.~\ref{GCEspectrum}) which suggests that photons that were
previously associated with Sgr A$^\ast$ are now being associated to
the GCE source. With the new diffuse and extended sources, Sgr
A$^\ast$ becomes nearly a power law with a log-parabola spectrum of
$N_0= (2.181\pm 0.082)\times 10^{-11}\rm\ MeV^{-1}\ cm^{-2}\ s^{-1}$,
$\alpha = 2.32\pm 0.032$, $\beta = 0.173\pm 0.020$, and a commensurate
reduction in flux to $(2.89\pm
0.18)\times10^{-7}\rm\ ph\ cm^{-2}\ s^{-1}$.

In Fig.~\ref{SgrAspectrum} we also show results of a banded SED fit
for Sgr A$^\ast$'s spectrum in the full 2FGL+2PS+I+MG+ND+GCE by
independently fitting the normalization of the Sgr A$^\ast$ flux while
fixing other sources within that energy band. This is similar to the
residual flux spectrum and provides a useful cross-check (see Appendix
for more details).

Note that our spectra for Sgr A$^\ast$ and the GCE source have a
spectral feature downturn and upturn at $E_\gamma\approx
1.3\rm\ GeV$. We find this feature in the full photon counts in the
ROI, and it is possible that this is an artifact of energy
identification in the Fermi tools at this energy.

Our best fit model for Sgr A$^\ast$ has implications for
interpretations of its gamma-ray emission. In the hadronic scenario,
the $\sim$GeV peak is associated with emission from diffusively
trapped protons. As the protons transition to rectilinear motion at
higher energies, they reproduce the flatter spectrum observed at
$\mathcal O\rm (TeV)$
energies~\cite{Chernyakova:2011zz,Linden:2012iv}.  In the context of
this scenario, the newly determined flatter spectrum near $\sim$1 GeV
implies changes to the diffusion parameters. For example, reasonable
reductions to the diffusion coefficient energy dependency and/or
diffusion coefficient normalization can generate such flatter spectra
~\cite{Chernyakova:2011zz}. Alternatively, reducing the Sgr A$^\ast$
activity duration, or reducing the proton diffusion region to smaller
than the saturation level of $3\rm\ pc$ as described in
Ref.~\cite{Linden:2012iv}, reduces the $\sim$GeV intensity and
generates a flatter spectrum.

When fitting in our full model with extended sources and the new
diffuse component, the 2FGL+2PS+I+MG+ND+GCE model, the emission
associated with the MG has a spectrum best fit with $N_0 = (1.68\pm
0.14)\times 10^{-9}\rm\ MeV^{-1}\ cm^{-2}\ s^{-1}\ sr^{-1}$, $\alpha =
1.487\pm 0.075$, $\beta = 0.297\pm 0.031$ for $E_b=300\rm\ GeV$. The
best fit spectra for the MG and ND components are shown in
Fig.~\ref{MG+ND_residspectrum}, along with the GCE spectrum for reference.

Reference \cite{YusefZadeh:2012nh} interpreted the gamma-ray emission
from the 20 cm correlated MG to be from bremsstrahlung of a
high-energy population of electrons on the molecular gas. However, our
new model xfits with additional sources reveal an intensity peaked at
energies of $\sim\!\!1\rm\ GeV$, which is slightly high. In
bremsstrahlung, typically half the $e^\pm$ energy is emitted; thus,
the gamma-ray spectrum follows the cosmic-ray $e^\pm$ spectrum. The
electron spectrum in turn is set by the dominant cooling or escape
processes. The bremsstrahlung energy loss time as $e^\pm$ traverse
pure hydrogen of number density $n$ is $t_{\rm brems} \approx 40 \,
(n/{\rm cm}^{-3})^{-1} \rm {\rm Myr}$, but since the ionization loss
time $t_{\rm ion} \approx 1380 \ E_{\rm GeV} (n/{\rm cm}^{-3})^{-1}
[\ln E_{\rm GeV} + 14.4]^{-1}$ dominates at low energies, the $e^\pm$
and gamma-ray spectra soften, yielding a peak at $\sim 400$ MeV,
independent of the target density.  On the other hand, the break could
result from a break in the cosmic ray (CR) electron spectrum. As
argued in Ref.~\cite{YusefZadeh:2012nh}, such an interpretation is
consistent with the observed radio emission in the GC region.

Based on the bremsstrahlung interpretation, information of the molecular 
gas density can be obtained. The MG and ND spectra above the peak 
imply a CR electron spectrum $dN/dE \propto E^{-p}$ with $p \sim 3$. 
The same CR electron population will synchrotron radiate in the radio 
with a spectrum $F_\nu \propto \nu^{-\alpha}$ and $\alpha = (p-1)/2 \sim1$. 
For a power-law CR electron population, the synchrotron radio and 
bremsstrahlung gamma emissions are related by, e.g., Eq.~(12) of 
Ref.~\cite{YusefZadeh:2012nh}. We adopt a magnetic field of 
$10 \, {\rm \mu G}$ in the GC region, which is within a factor of 2 of the range 
estimated from the CR ionization rate \cite{YusefZadeh:2012nh}, and implies
an electron of energy $E_e$ radiates $\sim 5 (B/10{\rm \mu G})(E_e/6 {\rm GeV})^2$ 
GHz radio and emits $\sim 3 (E_e/6 {\rm GeV})$ GeV gamma rays. 
Requiring that the observed radio at $5$ GHz towards the GC ($S_{5 {\rm GHz}} 
\sim 10^3$ Jy \cite{YusefZadeh:2012nh}) is not overpredicted, the MG and ND 
estimates imply a lower limit on the molecular gas density of  
$n_H \gtrsim 4 \, {\rm cm^{-3}} ({S_{5 {\rm GHz}}/1000 {\rm Jy}})^{-1}$. 

The emission associated with the new diffuse source for the full
model, the best fit log-parabola spectrum is $N_0 = (1.69\pm
0.39)\times 10^{-5}\rm\ MeV^{-1}\ cm^{-2}\ s{-1}\ sr^{-1}$, $\alpha =
0.95\pm 0.17$, $\beta = 0.308\pm 0.047$ for $E_b=100\rm\ MeV$. This is
essentially the same as the MG spectrum and this result likely
indicates the presence of molecular gas not captured by the Galactic
diffuse model and the MG template.

For the analysis with photons in the restricted $0.7-7$ GeV energy range, we did 
not detect the $\Gamma=-0.5$ ND source. Hence, we only show results 
for the E7 analysis without including the ND source, {\em i.e.}, 
E7-2FGL+2PS+nPS+MG+GCE. 
The MG spectrum in the E7 energy window has an index of almost -2.0 (with no 
significant variations), which is different from the fit using the full model. This is not
altogether surprising given the weight from lower energy photons in
constraining the MG spectrum in the full model. The differences may
also be due to degeneracies between GCE and MG in this restricted 
energy window given the similarity in their spectra at energies above 
about a GeV (see Fig.~\ref{MG+ND_residspectrum}).

In the full model, 2FGL+2PS+I+MG+ND+GCE, the emission associated with
the GCE source is best fit by log-parabola spectrum with $N_0 =
(1.20\pm 0.46)\times 10^{-12}\rm\ MeV^{-1}\ cm^{-2}\ s{-1}\ sr^{-1}$,
$\alpha = -4.28\pm 0.18$, $\beta = 0.959\pm 0.026$ for
$E_b=100\rm\ MeV$.  The GCE emission is almost equally well fit by 
a power law with an exponential cutoff 
$dN/dE = N_0 (E/E_0)^{-\gamma_c} \exp(-E/E_c)$
and the best fit spectral parameters are 
$\gamma_c = 0.45\pm0.21$, $E_c = 1.65\pm0.20\rm\ GeV$ and 
$N_0 = (1.03\pm0.56) \times 10^{-9}\rm\ MeV^{-1} cm^{-2}\ s^{-1}\ sr^{-1}$
for $E_0 = 100\rm\ MeV$.

One of the key features of the GCE excess is the striking similarity
to the $\rho^2$ spatial profile expected of annihilation signals. To
investigate this further we did two tests with the E7 data. First, for
the E7-2FGL+2PS+nPS+MG+GCE, we plotted the residual flux spectra in
different spatial regions and that is shown in
Fig.~\ref{GCEprofile}. It is clear that the excess is present
throughout the ROI and not just concentrated at the center. This is
partly why the GCE is robustly found in different analyses. We take
this one step further with a new model
E7-2FGL+2PS+nPS+MG+GCE(a)+GCE(b) where GCE(a) is GCE with pixels
outside a radius of $2.5^\circ$ zeroed out and GCE(b) = GCE - GCE(a)
is the complementary region with $\gamma=1.1$ in all cases.  We found
that there are fits that are statistically almost as good as the
E7-2FGL+2PS+nPS+MG+GCE ($\gamma = 1.1$) case but have different
spectra for the inner and outer parts. In particular, the best fit
peak in intensity for the outer part seems to be at somewhat larger
energy (but still between 2 and 3 GeV) . The $\Delta \ln \mathcal L$ is around
10 for these models compared to the E7-2FGL+2PS+nPS+MG+GCE ($\gamma =
1.1$) case and that is not significant enough to claim deviations from
our baseline model with GCE.
 
What the above does bring up is the possibility that the fit can
accommodate more than one diffuse component as part of the
GCE---perhaps due to MSPs and dark matter.  This exciting possibility
deserves further study and we suggest that it should be considered
equally as likely as the pure dark matter hypothesis since the best fit
spectrum from dark matter annihilation is very similar to the MSP
spectrum~\cite{Abazajian:2010zy}. To illustrate this point, we show a
plot of the GCE spectra from our full model compared to the spectra of
eight globular clusters that were observed with Fermi LAT. We have focused
in on the region around a GeV and higher since that is where we are
(comparatively) more confident in our background modeling. We have
also normalized all the spectra by their fluxes for $E>2$ GeV to make
the comparison easier. The similarity of the GCE excess with the 
spectra from globular clusters is readily apparent.

\subsection{Dark matter interpretation}
\label{dmsection}

\begin{figure}[t]
\begin{center}
\includegraphics[width=3.0truein]{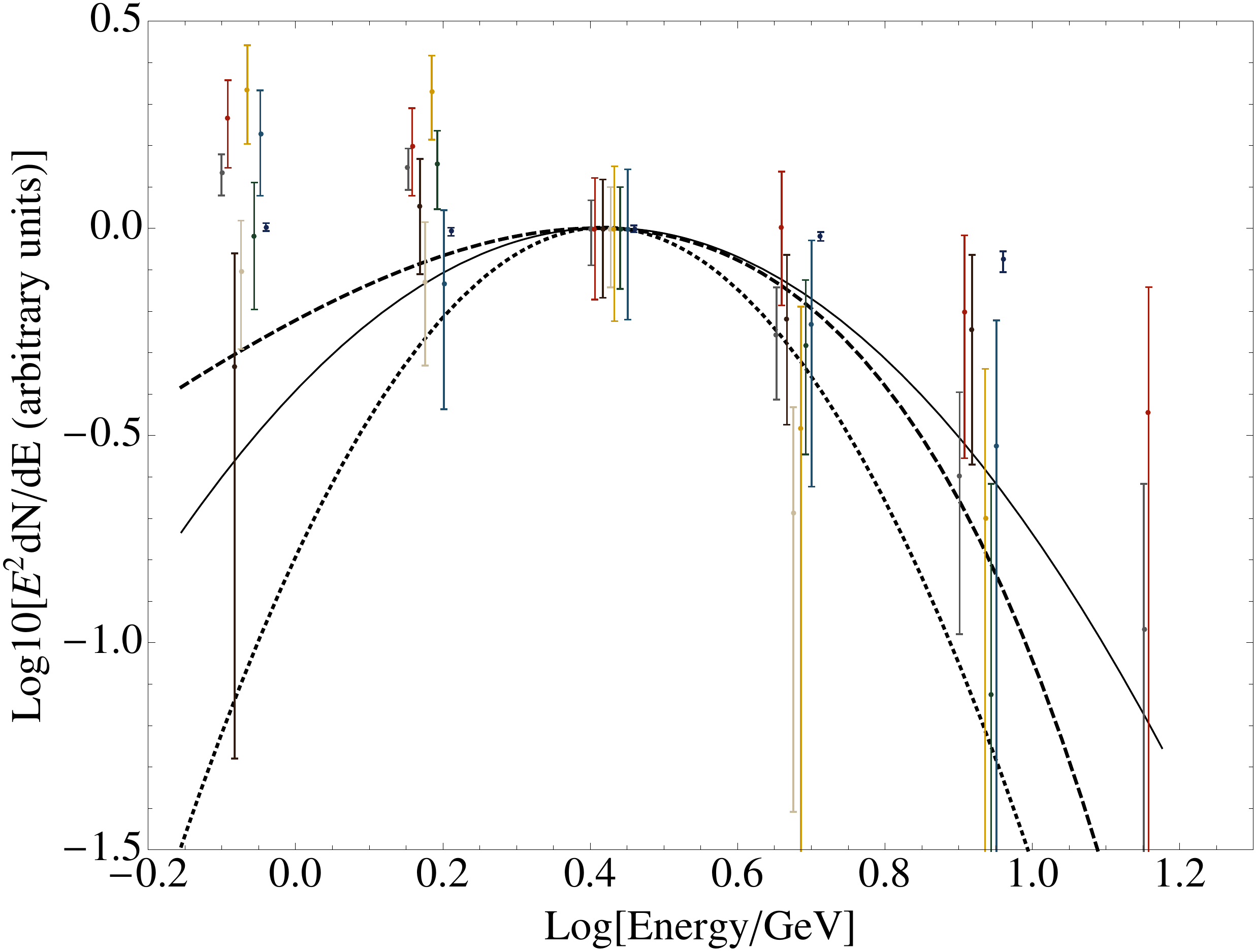}
\end{center}
\caption{Here we compare the flux spectra of the best fit GCE source
  with the flux spectra from eight globular clusters detected by Fermi
  LAT (47 Tuc, $\omega$ Cen, M62, NGC 6388, Terzan 5, NGC 6440, M28,
  NGC 6652). The three best fit GCE spectra shown are from the full
  model with a power-law exponential cutoff spectrum (solid), from the
  full model with a log-parabola spectrum (dashed) and from the 0.7--7
  GeV analysis with a log-parabola spectrum (dotted). All the spectra
  are normalized by their fluxes for energies greater than 2 GeV.
\label{flux_compare}}
\vskip -0.3cm
\end{figure}

\begin{table}[t]
\caption{Flux, in units of $10^{-7} \rm\ ph\ cm^{-2}\ s^{-1}$ within 0.2
  - 300 GeV, in the $7^\circ\times 7^\circ$ ROI and $\ts =2\Delta
  \ln(\mathcal L)$ values for several sources in the GC region for our
  full 2FGL+2PS+I+MG+ND model. The TS values are determined with
  reoptimization of the respective models with the same morphological
  parameters $\gamma$ and $\Gamma$. We leave the TS value for the
  Galactic diffuse case as an approximation due to its very high
  significance.}
\label{fluxtable}
\begin{ruledtabular}
\begin{tabular}{l|rr}
 Source Name & Flux   & $\ts$ \\
 \hline \\
 2FGL J1745.6-2858 (Sgr A*) & $2.89\pm 0.18$ & 789.6  \\
 2FGL J1747.3-2825c (Sgr B) & $0.573\pm 0.098$  & 179.8 \\
 2FGL J1746.6-2851c (the Arc) & $0.773\pm 0.182$  & 67.1 \\
 2FGL J1748.6-2913 & $0.361\pm 0.082$  &  90.3 \\
 MG & $7.29\pm 0.52$ & 185.7 \\
 GCE $\gamma=1.1$ & $1.08\pm 0.10$ & 170.7 \\ 
 ND $\Gamma=-0.5$ & $2.99\pm 0.38$ & 73.5 \\
 Galactic diffuse & $34.8\pm 0.46$ & $\gtrsim 10^4$ \\
\end{tabular}
\end{ruledtabular}
\end{table}

\begin{figure*}[t]
\begin{center}
\includegraphics[width=3.4truein]{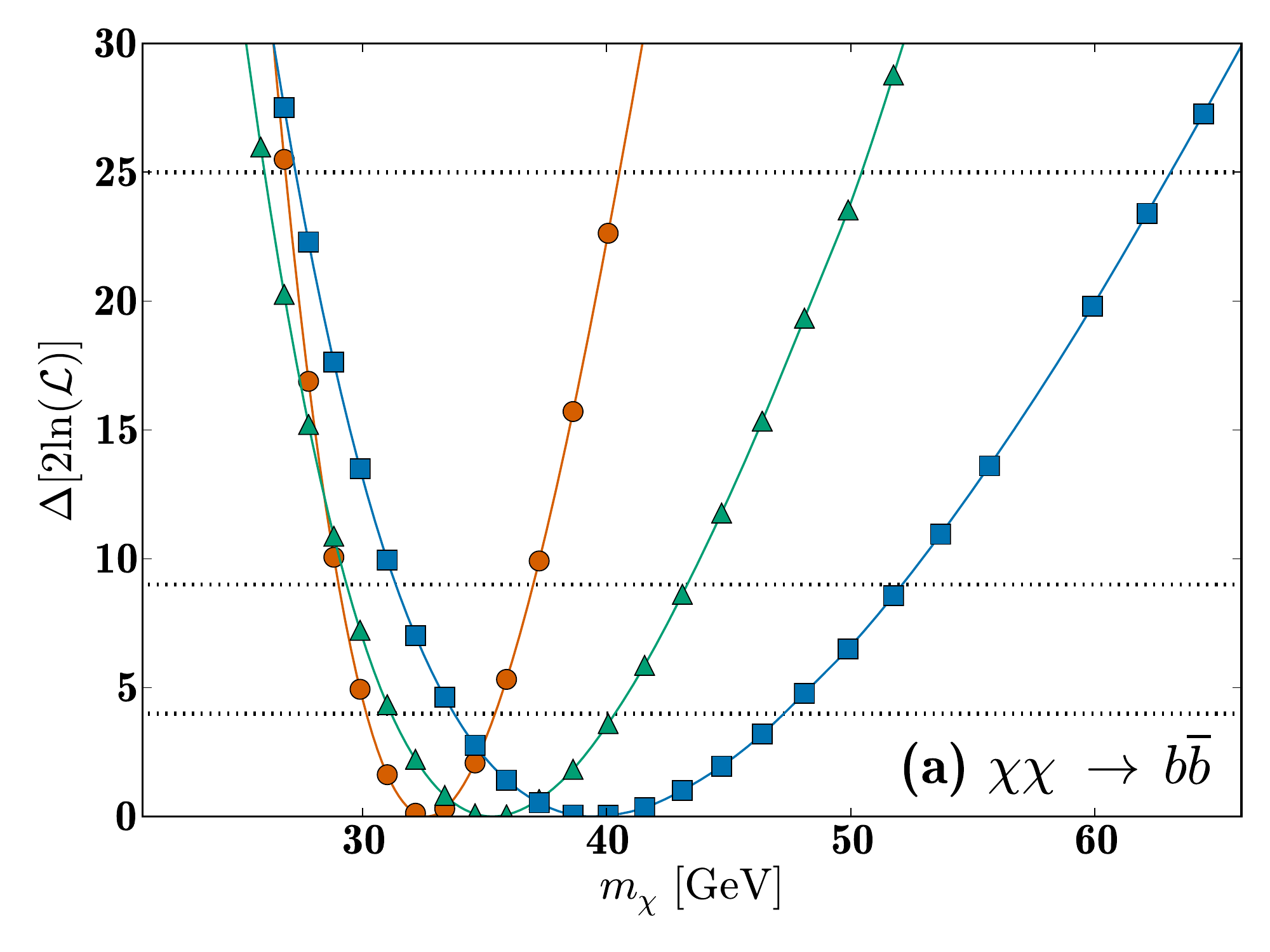}
\includegraphics[width=3.4truein]{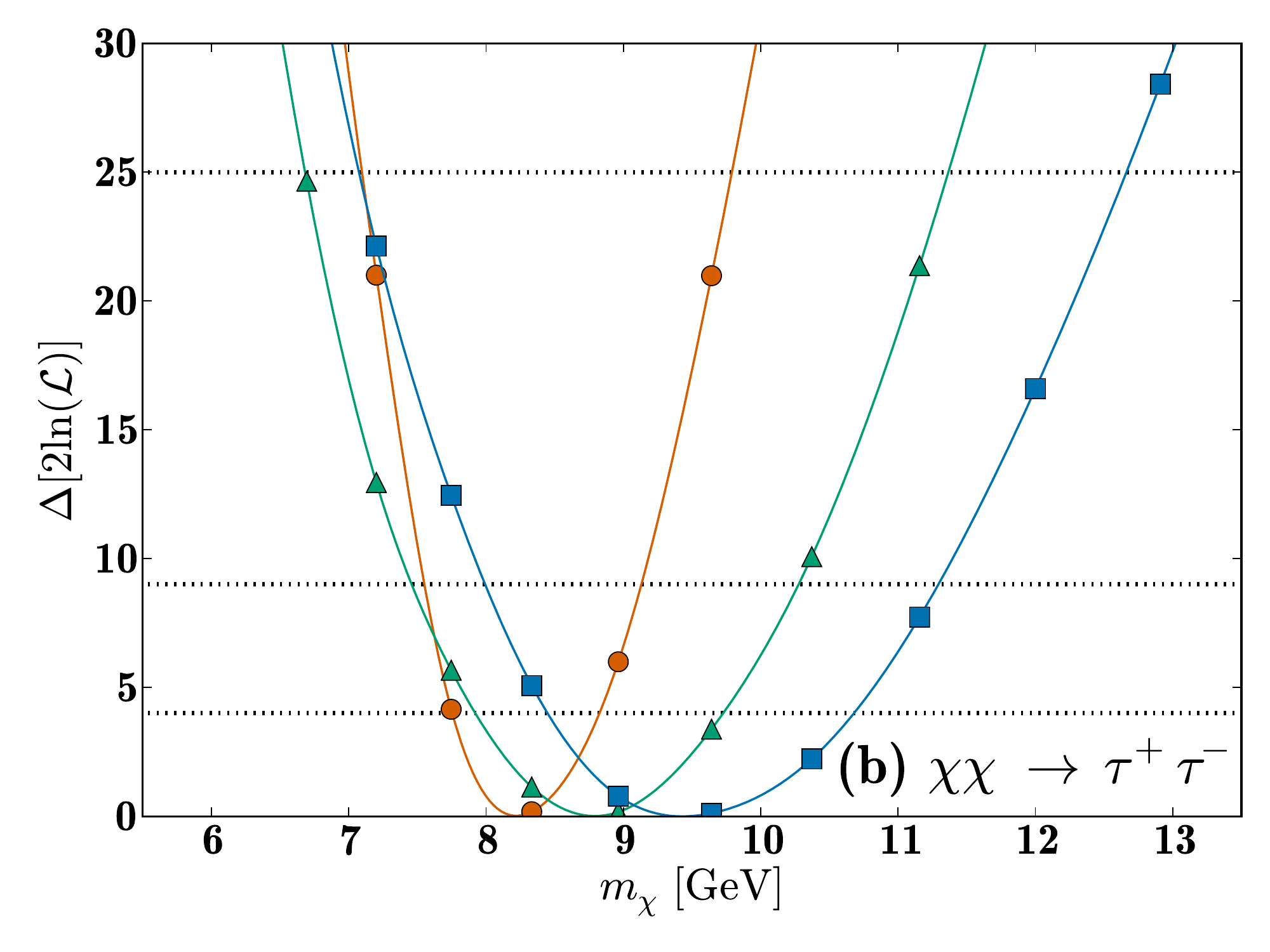}\hskip -0.2 in
\end{center}
\caption{Shown are the $2\Delta \ln(\mathcal L)$ for the best fit dark
  matter particle masses for (a) pure $b\bar b$ and (b) pure
  $\tau^+\tau^-$ annihilation channels, for several astrophysical
  model cases studied when varying all sources on the GC ROI. In both
  panels, the cases for 2FGL+2PS+GCE show the exact particle mass runs
  in orange circles, 2FGL+2PS+I+MG+GCE case in green triangles and the
  full best fit model 2FGL+2PS+I+MG+ND+GCE case in blue squares. Fourth
  order spline interpolations are shown as lines for each case, which
  are used to find the minima and limits. For the full 2FGL+2PS+I+MG+ND+GCE
  model, the $b\bar b$ and $\tau^+\tau^-$ are equivalent in their
  goodness of fit, and there is no evidence for a mixed
  channel. The horizontal lines are for $2, 3$ and $5\sigma$
  limits.  \label{massfit}}
\end{figure*}

When interpreting the GCE source as originating in dark matter
annihilation, we found that the best fit mass for annihilation into $b
\overline{b}$ was $31.4^{+1.4}_{-1.3}$, $35.3^{+2.4}_{-2.2}$, and
$39.4^{+3.7}_{-2.9}$ GeV for the 2FGL+2PS+GCE, 2FGL+2PS+I+MG+GCE, and
2FGL+2PS+I+MG+ND+GCE models, respectively. The amplitude for annihilation
rate $\langle \sigma v\rangle_{b\bar b}$ for the full model
2FGL+2PS+I+MG+ND+GCE is $(5.1 \pm 2.4)\times
10^{-26}\rm\ cm^3\ s^{-1}$.  For annihilation into $\tau^{+}
\tau^{-}$, the best fit masses were $8.21^{+0.30}_{-0.24}$,
$8.79^{+0.44}_{-0.42}$ and $9.43^{+0.63}_{-0.52}$ GeV for the
2FGL+2PS+GCE, 2FGL+2PS+I+MG+GCE, and 2FGL+2PS+I+MG+ND+GCE models,
respectively.  The amplitude for annihilation rate $\langle \sigma
v\rangle_{\tau^{+} \tau^{-}}$ for the full case 2FGL+2PS+I+MG+ND+GCE
is $(0.51\pm 0.24)\times 10^{-26}\rm\ cm^3\ s^{-1}$.\footnote{The
  errors on $\langle \sigma v\rangle$ are dominated by the uncertainty
  in the local dark matter density, which we adopt as $\rho_\odot =
  0.3\pm 0.1\rm\ GeV\ cm^{-3}$~\cite{Zhang:2012rsb}.}  These mass fit
$2\Delta \ln(\mathcal L)$ curves are shown in Fig. \ref{massfit}.

\begin{figure}[t!]
\begin{center}
\includegraphics[width=3.4truein]{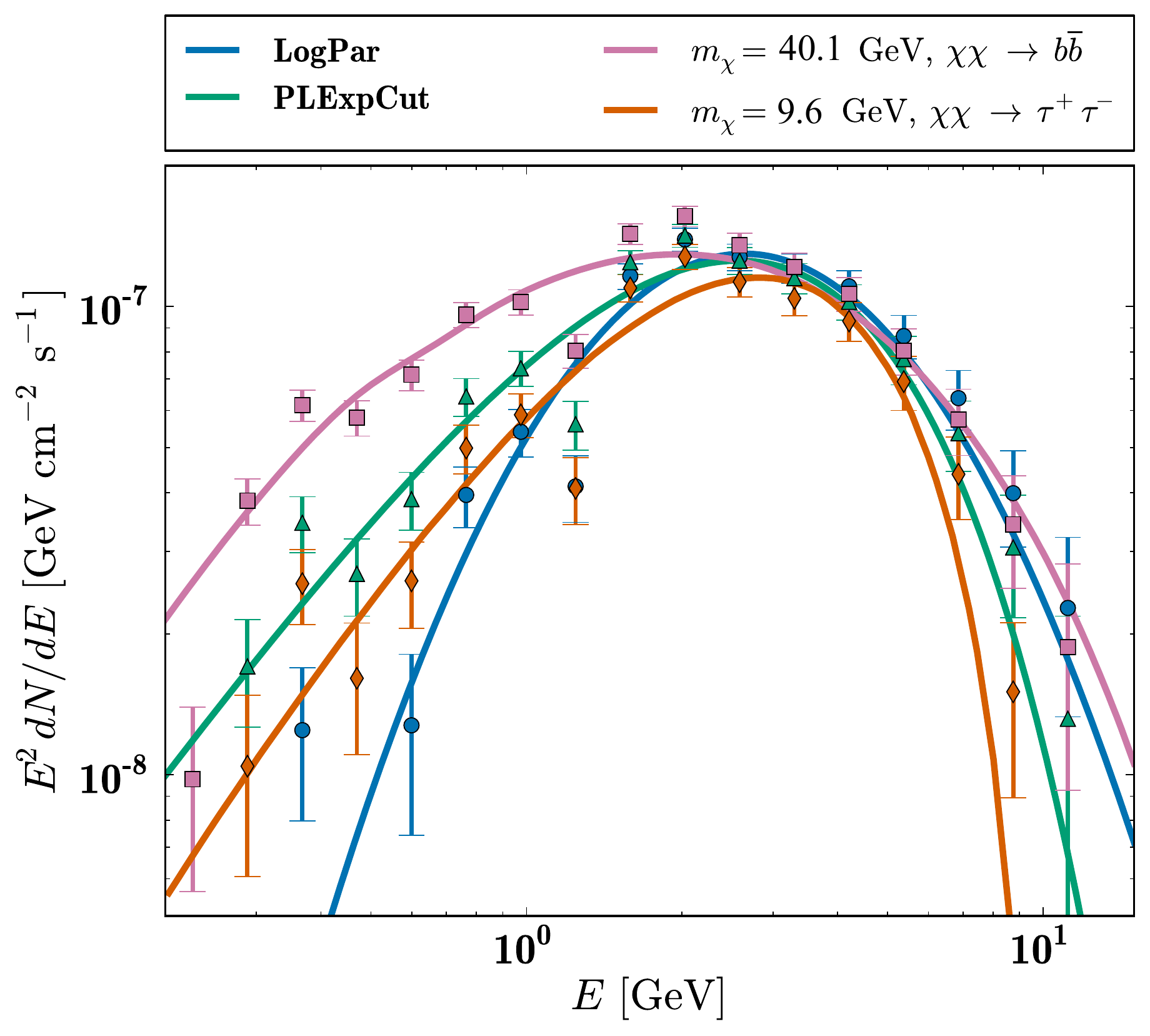}
\end{center}
\caption{Shown are the systematic and statistical uncertainties in
  determining the GCE source spectrum. The errors represent the
  SED-normalization statistical uncertainty within an energy band,
  while the several cases represent the inherent systematic
  uncertainty present in the adoption the GCE source's spectral
  form. \label{systematicsspectrum}}
\vskip -0.3cm
\end{figure}

When using the 2FGL+2PS+I+MG model, the $b$-quark channel is preferred
over $\tau$ leptons by a $\Delta \ln(\mathcal L) \approx 17.9$. This
is consistent with recent results applying the 20 cm radio map as well
as Galactic ridge template models to dark matter annihilation models
\cite{Macias:2013vya}, which find a preference for the $b$-quark
annihilation channel.  As can be seen in Figs.~\ref{GCEspectrum} and
\ref{systematicsspectrum}, the steepness of the rise of the spectrum
is highly diffuse-emission model and GCE-spectral model dependent, and
it is therefore problematic to draw conclusions on the nature of the
emission from the residual spectra and rise shapes of SED spectra
alone, as is done, e.g., in
Refs.~\cite{Hooper:2010mq,Hooper:2011ti}. These large variations in
best fit spectra (specifically below about GeV) are indicative of
degeneracies that can only be accounted for in a full likelihood
spatial and spectral analysis of the type performed here and in
Ref.~\cite{Macias:2013vya}.

In the case of mixed channels (arbitrary branching ratio into $b
\overline{b}$ and $\tau^{+} \tau^{-}$) in the full model,
2FGL+2PS+I+MG+ND+GCE, we find no preference for mixed channels, with
the likelihood profile having a minimum at full $b$-quark channel
annihilation at higher $m_\chi\approx 30-40\rm\ GeV$ and annihilation
into $\tau$ leptons at lower masses $m_\chi\approx 10\rm\ GeV$, with
these two minima separated only by $\Delta \ln(\mathcal L)=0.8$.  If
we do not include the molecular gas contribution, then the preferred
dark matter masses shift to lower values.  

Importantly, bremsstrahlung effects of the annihilation products can
appreciably modify the gamma-ray spectra~\cite{Cirelli:2013mqa}. In
particular, the work in Ref. \cite{Cirelli:2013mqa} found that the
$\tau^{+} \tau^{-}$ channel is softened, or, less steep at low
energies, under standard assumptions for the gas density and magnetic
fields in the GC.

To test the magnitude of the effects of bremsstrahlung of final state
particles in the astrophysical environment of the GC, we utilize the
following approximation of the effects. We apply the bremsstrahlung
spectra for the ``realistic gas density'' for the
$m_\chi=25{\rm\ GeV}$ $b \overline{b}$-channel and
$m_\chi=20{\rm\ GeV}$ $\tau^{+} \tau^{-}$-channel cases in Fig.\ 4 of
Ref.~\cite{Cirelli:2013mqa} as the magnitude of the effect for all
particle masses of interest. We scale the bremsstrahlung photon
spectra energies with the particle masses proportionally with the
prompt spectra over our particle mass range. We then rederive the
best fit particle mass determinations. This method is an approximation
of the bremsstrahlung effects, but provides an order-of-magnitude
estimate of the modification of gamma-ray spectra due to particle
bremsstrahlung in the annihilation cascade.  When adding the
bremsstrahlung photons in the manner described, we find that the
best fit particle masses are, for the $b \overline{b}$ channel,
$m_\chi=40.9^{+3.6}_{-3.4}\rm\ GeV$, and for the $\tau^{+}
\tau^{-}$-channel, $m_\chi=10.17^{+0.54}_{-0.59}\rm\ GeV$. The larger
best fit masses reflect the softening of the spectra that allows more
massive particles to fit the observed photon spectrum. Because the effect
is relatively small, this shift is subsumed in the systematic errors
in Eqs.~\eqref{bmass} and \eqref{taumass} below, which are dominated by
diffuse model uncertainties. However, it is notable that with the
bremsstrahlung spectral modification, we find that the $\tau^{+}
\tau^{-}$ channel is preferred by $\Delta \ln(\mathcal L)=4.5$, which
is statistically significant at approximately $\sim\!\! 3\sigma$. More
detailed work on the particle bremsstrahlung is warranted, but beyond
the scope of this paper.

The statistical error on the dark matter particle mass producing the
signal is quite small in these cases, at better than $10\%$ in all
cases. However, the systematic error associated with uncertainties in
the astrophysical diffuse models, present in particular with true
fractional MG contribution along the line of sight, render the
systematic uncertainty relatively large, at about 20\%. Therefore, our
determination of the dark matter particle mass and annihilation rate
in the pure $b \overline{b}$ channel is
\begin{eqnarray}
m_\chi &=& 39.4\left(^{+3.7}_{-2.9}\rm\ stat.\right)\left(\pm
7.9\rm\ sys.\right)\rm\ GeV  \label{bmass} \cr
\langle \sigma v\rangle_{b\bar b} &=& (5.1\pm 2.4)\times 10^{-26}\rm\ cm^3\ s^{-1}\, ,
\end{eqnarray}
where the best fit value is determined by the full model,
2FGL+2PS+I+MG+ND+GCE. The annihilation rate is below the most stringent
constraint on this region, from the four year combined dwarf analysis,
with an upper limit requiring $\langle\sigma v\rangle_{b\bar b}
\lesssim 6.5\times 10^{-26}\rm\ cm^3\ s^{-1}$ (95\% C.L.) \cite{Ackermann:2013yva}.  

Note that there are significant constraints on the annihilation
through specific interaction operators at comparable rates from dark matter
searches at the Large Hadron
Collider~\cite{Chatrchyan:2012me,ATLAS:2012ky,Goodman:2010ku}. In
particular, annihilation into quarks at our best-fit $m_\chi$ is
constrained by ATLAS \cite{ATLAS:2012ky} to be $\langle\sigma
v\rangle_{\tau^{+} \tau^{-}} \lesssim 2 (40) \times
10^{-26}\rm\ cm^3\ s^{-1}$ (95\% CL) for axial-vector (vector)
interaction couplings.

In the case of a pure $\tau^{+} \tau^{-}$ channel we find
\begin{eqnarray}
\label{taumass}
m_\chi &=& 9.43\left(^{+0.63}_{-0.52}\rm\ stat.\right)\left(\pm 1.2\rm\ sys.\right)\ {\rm GeV} \cr
\langle \sigma v\rangle_{\tau^{+} \tau^{-}} &=& (0.51\pm 0.24)\times 10^{-26}\rm\ cm^3\ s^{-1}\, ,
\end{eqnarray}
where the best-fit value is again determined by the full model,
2FGL+2PS+I+MG+ND+GCE. The annihilation rate in this channel is also
below the most stringent constraint on this region, from the 4 year
combined dwarf analysis, with an upper limit requiring $\langle\sigma
v\rangle_{\tau^{+} \tau^{-}} \lesssim 2.3\times
10^{-26}\rm\ cm^3\ s^{-1}$ (95\% CL) \cite{Ackermann:2013yva}. As
discussed above, our determined uncertainties in $\langle \sigma
v\rangle$ are dominated by the local dark matter density uncertainty.
There are systematic uncertainties on the annihilation rates in
Eqs.~\eqref{bmass} and \eqref{taumass} due to the diffuse model and dark
matter profile $\gamma$ uncertainties, but they are smaller than the
uncertainties due to the local dark matter density.

Interpreting the GCE emission in dark matter models beyond the single
channel cases we present here requires significant care. The nature of
the GCE source and photons associated with the source depends on the
underlying assumption of the spectrum and morphology of the dark
matter GCE source, as well as the modeling of the other diffuse and
point sources in the region, as discussed above and shown in
Fig.~\ref{GCEspectrum}. To illustrate, we show the GCE spectra for our
full model for several spectral model cases in
Fig.~\ref{systematicsspectrum}. Here, we fit the spectral energy
distribution (SED) of the GCE source independently in energy bins
across the energy range of interest, while keeping the other sources
fixed in that energy bin. This provides an estimate of the statistical
uncertainty of the GCE source spectrum including covariance with other
source fluxes. We refit the SED with this method for the log-parabola,
power law with exponential cutoff, as well as the $b$-quark and
$\tau$-annihilation channels. It is clear from
Fig.~\ref{systematicsspectrum} that the derived nature of the source
spectrum depends on the assumed spectrum. Though still approximate,
the best estimate of the GCE spectrum, including its overall
statistical and systematic uncertainty, would be the full range of
errors between the upper-most and lower-most points' errors in
Fig.~\ref{systematicsspectrum}.

\subsection{Astrophysical interpretations \& limits on dark matter contribution}
\label{limitssection}

\begin{figure*}[t]
\begin{center}
\includegraphics[width=3.4truein]{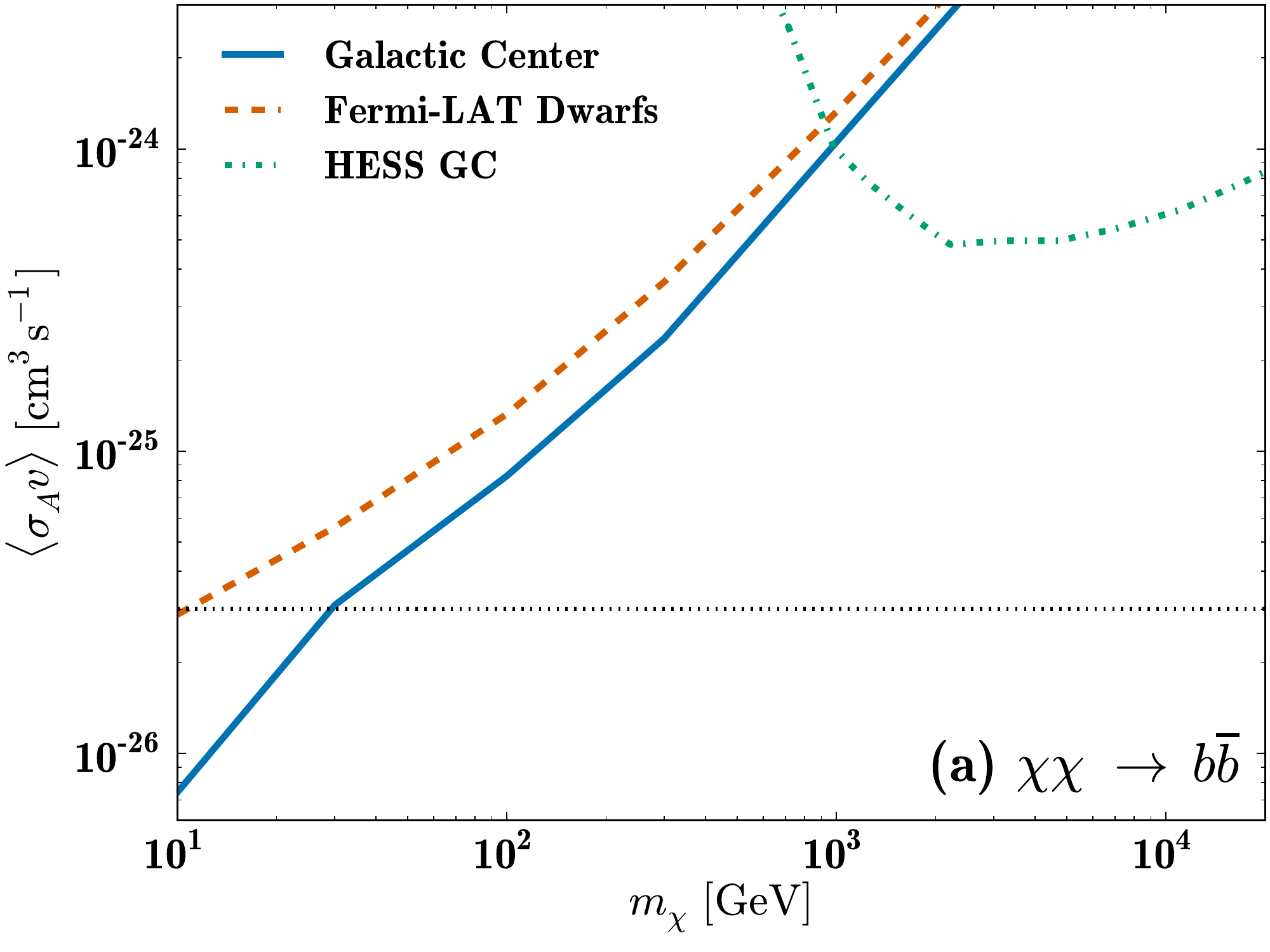}
\includegraphics[width=3.4truein]{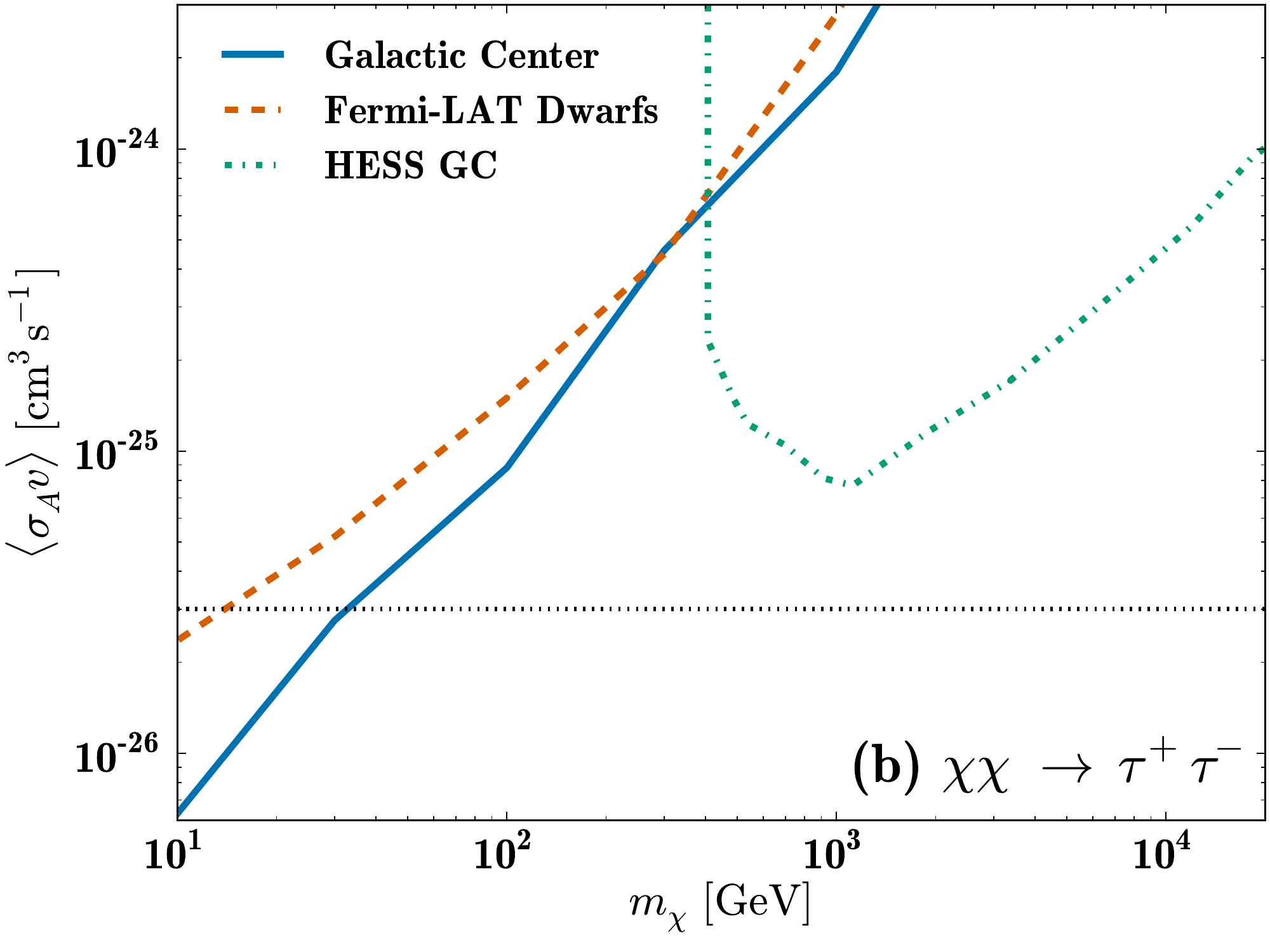}\\
\includegraphics[width=3.4truein]{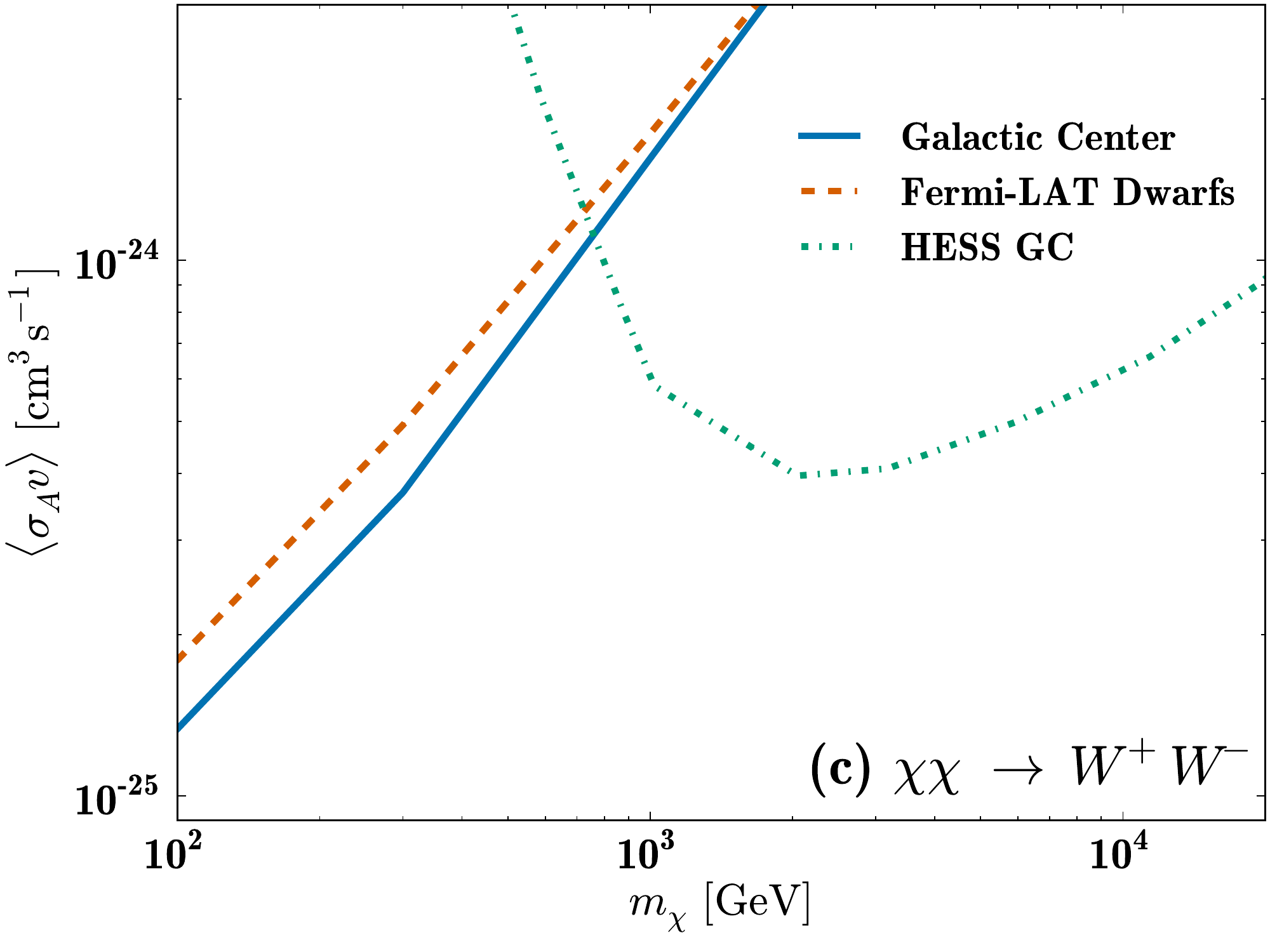}
\includegraphics[width=3.4truein]{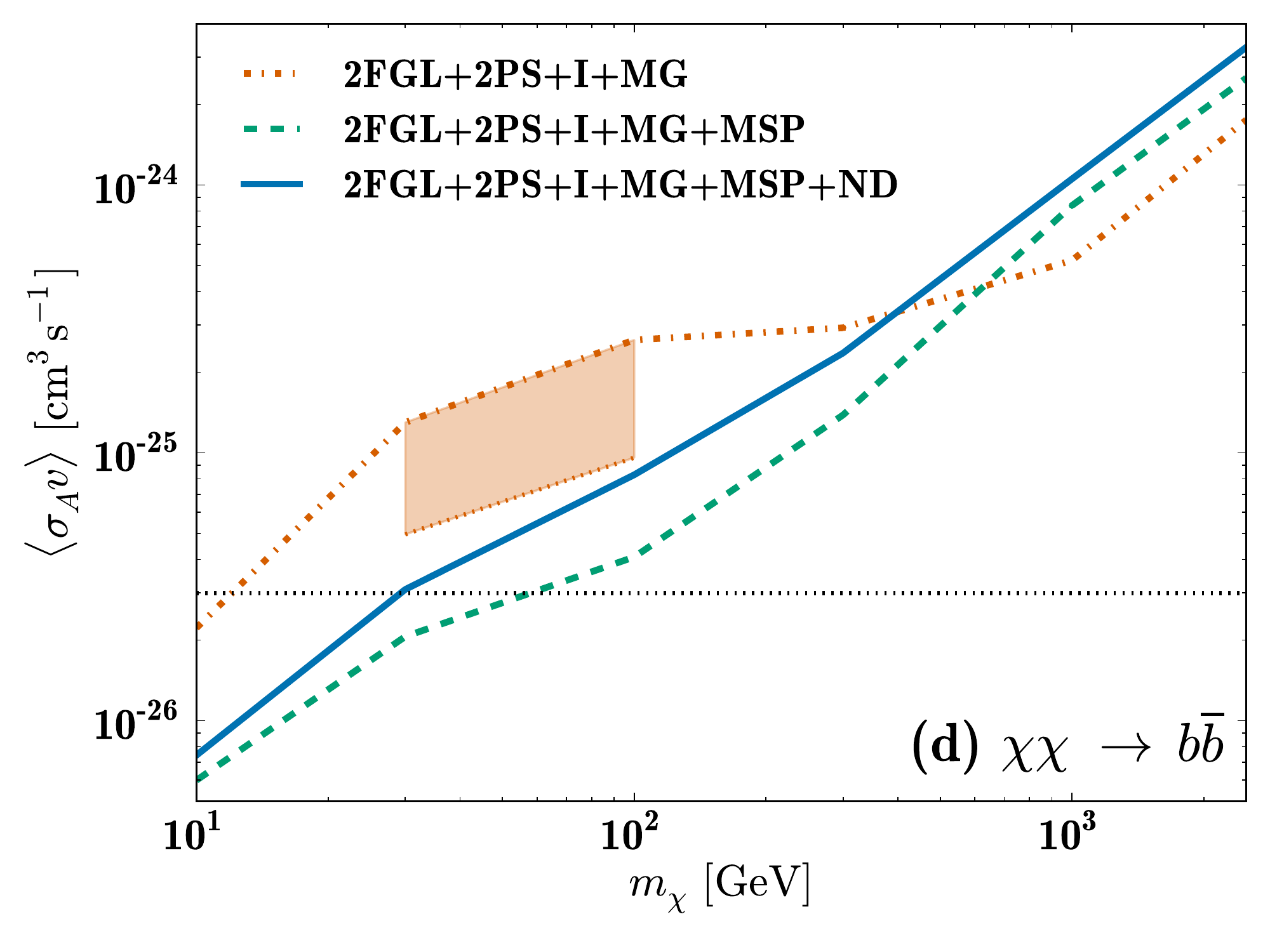}
\end{center}
\caption{Shown are limits on several channels when assuming that the
  new extended source is associated with MSP or other astrophysical
  emission in the models we study, for (a) the $b\bar b$, (b)
  $\tau^+\tau^-$, and (c) $W^+W^-$, in comparison with combined dwarf
  galaxy limits \cite{Ackermann:2013yva} and limits from HESS
  observations toward the Milky Way GC~\cite{Abazajian:2011ak}. In (d)
  we show the strong model dependence of the limits, with the adopted
  full model limits being 2FGL+2PS+I+MG+MSP+ND solid (blue). The
  shaded box is for the case of 2FGL+2PS+MG, where there is the
  detection.  \label{limits}}
\end{figure*}

There were significant detections of an extended source consistent
with a dark matter interpretation into the quark channel in all of our
models. However, as discussed in the introduction and in previous
studies, this emission is also consistent with a population of MSPs as
shown by the comparison of the spectra in Fig.~\ref{flux_compare}.  To
estimate the required MSP population within the ROI, we use 47 Tuc as
a reference. As we have seen previously, the flux estimates of the GCE
source have large systematic uncertainties below about 2 GeV. The
spectrum of the GCE is also more consistent with those of globular
clusters (including 47 Tuc) above this energy. So we choose to compare
the fluxes at $E>2\ {\rm GeV}$.  If 47 Tuc were at the GC its flux
above 2 GeV would be $3 \times 10^{-10}\ {\rm cm}^{-2} {\rm
  s}^{-1}$. The current estimate for the number of MSPs in 47 Tuc is
around 30. We use this to estimate the flux per MSP contributing to
the GCE to be $10^{-11} {\rm cm}^{-2} {\rm s}^{-1}$. The total flux
for the best power law with exponential cutoff spectrum is $4.8\times
10^{-8}\ {\rm cm}^{-2} {\rm s}^{-1}$, which implies about 4800 MSPs
are required within the ROI, while the same calculation for the
log-parabola spectrum from the full model yields 3700 MSPs within the
ROI.

Consistent with previous work, when we included a dark matter source
in addition to the MSP source, there was no significant dark matter
detection, because we assumed the spatial morphologies to be the same
\cite{Abazajian:2012pn} and since the log-parabola spectrum is
sufficiently flexible.  If we assume that all of the GCE emission is
astrophysical (e.g., unresolved MSPs), we can place limits on the
annihilation cross section for a potential WIMP contribution. We find
that this limit is highly dependent on which model components we
include. The various limits for annihilation into $b \overline{b}$ and
their dependence on three different models can be seen in
Fig. \ref{limits}.

We derive the 95$\%$ C.L. limits on the dark matter
annihilation cross section given each of these astrophysical models by
increasing the flux from the best fit value for the dark matter source
and then refitting all significantly detected parameters in the ROI
until $2 \Delta \ln(\mathcal L)=2.71$ for the one-sided confidence
level. This is done for the $b \overline{b}$ and $\tau^{+} \tau^{-}$
channels for masses 10, 30, 100, 300, 1000, and 2500 GeV, and for
the $W^{+} W^{-}$ channel for masses 100, 300, 1000, and 2500 GeV. We use
only photons from 700 MeV to 300 GeV as this range was found to
provide a more stringent limit.

For our adopted shown limits, we use our full 2FGL+2PS+I+MG+ND+GCE
model, i.e., including the two additional point sources, the new
isotropic component, the MG template, $\gamma=1.1$ MSP template, a
$\gamma=1.0$ DM template, and the new diffuse component with
$\Gamma=-0.5$. These limits are shown in
Figs.~\ref{limits}(a)-\ref{limits}(c) for annihilation in $b
\overline{b}$, $\tau^{+} \tau^{-}$, and $W^{+} W^{-}$, and are
slightly more stringent than the four year Fermi stacked dwarf limits
\cite{Ackermann:2013yva}. We also show, for comparison, the limits
from High Energy Stereoscopic System (HESS) observations toward the Milky Way
GC~\cite{Abramowski:2011hc,Abazajian:2011ak}.  Note, however, the GC
limits are highly dependent on the adopted diffuse-emission models, as
shown in Fig.~\ref{limits}(d). Therefore, though the GC DM limits are
stringent, they are not robust to underlying model assumptions,
contrary to some previous claims~\cite{Hooper:2012sr}.

\section{Conclusions}
\label{conclusions}

We have presented the results of a large set of analyses of the nature
of point source, diffuse and extended source gamma-ray emission toward
the Milky Way's Galactic Center as observed by the Fermi LAT. We have
included all known point sources toward the GC as well as a template
of the molecular gas based on radio emission. In all cases, we find a
highly statistically significant robust detection of an extended
source consistent with dark matter annihilation and/or a population of
millisecond pulsars in the GC. However, the detailed spectrum of this
extended source depends strongly on the background (diffuse source)
models.

The spectrum of the source associated with Sgr A$^\ast$ is less steep
than in previous work, owing to the new extended and diffuse
sources. In the case of a dark matter annihilation interpretation of
the GC extended source, the particle mass is very precisely determined
given an annihilation channel, though systematic uncertainties in the
diffuse emission introduce significant systematic uncertainties.  The
$b$-quark or $\tau$-lepton channels are almost equally preferred, but
with different particle masses.  For annihilation into $b$ quarks we
find $m_\chi=39.4\left(^{+3.7}_{-2.9}\rm\ stat.\right)\left(\pm
7.9\rm\ sys.\right)\rm\ GeV$, $\langle \sigma v\rangle_{bb} = (5.1\pm
2.4)\times 10^{-26}\rm\ cm^3\ s^{-1}$. For the $\tau^{+} \tau^{-}$
channel we find $m_\chi =
9.43\left(^{+0.63}_{-0.52}\rm\ stat.\right)(\pm 1.2\rm\ sys.)\ GeV$,
$\langle \sigma v\rangle_{\tau^{+} \tau^{-}} = (0.51\pm 0.24)\times
10^{-26}\rm\ cm^3\ s^{-1}$. These annihilation rates are lower than,
but close to the annihilation rates that are excluded by combined
dwarf galaxy analyses~\cite{Ackermann:2013yva} and collider
searches~\cite{ATLAS:2012ky}. Future combined dwarf galaxy analyses
may be sensitive to this parameter
space~\cite{KoushiappasAspen2013,DrlicaWagner:2013,He:2013jza}.  Once
confirmed, measurements of the isotropic extragalactic background can
yield further information on, e.g., the smallest halo mass
\cite{Ng:2013xha}.

It has been pointed out that bremsstrahlung will modify the gamma-ray
spectra appreciably \cite{Cirelli:2013mqa}, and our tests find that
they increase the inferred particle masses in the $b \overline{b}$ or
$\tau^{+} \tau^{-}$ channels.  While the extended source is robustly
detected, we caution that the shape of the rise and fall of the
spectrum ($E^2dN/dE$), as shown in Figs.~\ref{GCEspectrum} and
\ref{systematicsspectrum}, is highly model dependent.

When interpreting all of the GCE emission as astrophysical, we find
stringent limits on dark matter annihilation, but they are highly
model dependent. In this sense, the combined dwarf limits are still
the most robust.

To explain the diffuse GCE emission with unresolved MSPs, we estimated
(using the gamma rays from 47 Tuc as a reference) that there need to
be about 3000 to 5000 MSPs within the ROI (1 kpc by 1 kpc box towards
the GC). This is a large number compared to the typical number of MSPs
in globular clusters but the total stellar content is also much larger
in this region. We have also highlighted the possibility that multiple
sources may contribute to the GCE.

While we have characterized some of the systematic uncertainty
associated with modeling of the diffuse background, we emphasize that
our treatment is far from exhaustive. Further multiwavelength study of
the Milky Way's Galactic Center is essential to understanding the
nature of the numerous sources in this highly dense astrophysical
region. Even so, the detection of the GCE source is fairly robust to
differences in the background modeling, and though the extended
emission in gamma rays studied here is consistent with a pure
astrophysics interpretation, the extended emission's consistency in
morphology, spectrum and flux with a dark matter annihilation
interpretation remains extremely intriguing.

\appendix*

\section{Residual flux and error}
The plots in this paper show both the residual flux and an alternate
estimate of the spectral energy distribution. We summarize the methods
to create them both here.  The residual flux in some energy bin
$\alpha$ is
\begin{eqnarray}
r_\alpha = \sum_\beta \frac{(n_{\alpha\beta} - b_{\alpha\beta})}{\epsilon_{\alpha\beta}}\,,
\end{eqnarray} 
where $n_{\alpha\beta}$ and $b_{\alpha\beta} $ are the total counts
and the background model count (all sources minus the source of
interest), respectively. The sum is over all spatial bins within the
ROI or part of ROI, as desired and $\epsilon$ is the exposure. The
Poisson error on this flux is given by,
\begin{eqnarray}
\delta\! r_\alpha^2 = \sum_\beta \frac{m_{\alpha\beta}}{\epsilon_{\alpha\beta}^2}
\end{eqnarray}

An alternate way to estimate the SED is to fix the background ($b$)
and maximize the likelihood in each energy bin for the amplitude of
the source of interest. We note that this SED estimate does not 
account for the correlations between GCE and other source
parameters but it is the quantity most directly comparable
to the residual flux. This likelihood (up to a constant) is
\begin{equation}
\ln {\mathcal L}_\alpha = -\sum_\beta m_{\alpha\beta} + \sum_\beta n_{\alpha\beta} \ln(m_{\alpha\beta})
\end{equation}
Writing $m_{\alpha\beta} = b_{\alpha\beta} + a_\alpha s_{\alpha\beta}
$ where $s$ labels the counts for the source of interest, the maximum
likelihood estimate of $a_\alpha$ and the error on $a_\alpha$ are
given by
\begin{eqnarray}
\sum_\beta (n_{\alpha\beta}/m_{\alpha\beta}-1)s_{\alpha\beta} = 0\nonumber\\
\delta a_\alpha^{-2} = \sum_\beta n_{\alpha\beta} s_{\alpha\beta}^2/m_{\alpha\beta}^2 \nonumber
\end{eqnarray}
The SED estimate is $(a_\alpha \pm \delta\! a_\alpha) \sum_\beta s_{\alpha\beta}/\epsilon_{\alpha\beta}$.
The SED estimate and residual flux values generally agree with each other. 

\acknowledgments We thank Marcus Ackermann, Theresa Brandt, Roland
Crocker, Chris Gordon, Dan Hooper, Tim Linden, and Tracy Slatyer for
useful discussions. We thank Farhad Yusef-Zadeh for providing the 20
cm radio maps, and Tim Linden for the Sgr A$^\ast$ emission model in
Fig.~\ref{SgrAspectrum}.  K.N.A. and N.C. are partially supported by
NSF CAREER Grant No. PHY-11-59224, and S.H. by a JSPS fellowship for
research abroad.

\bibliography{master}
\end{document}